%% file: anish.tex
\documentclass[12pt]{article}

\usepackage{amsmath,amsthm, amsfonts, amssymb, amsxtra,amsopn}
\usepackage{graphicx}
\usepackage{multirow}
\usepackage{algorithmic}
\usepackage{colortbl}
\usepackage{pgfplots}
\usepackage{pgfplotstable}
\usepackage{adjustbox}

\usepackage{subcaption}

\usepackage[width=0.94\textwidth,font=small,labelfont=bf]{caption}

\PassOptionsToPackage{hyphens}{url}

\usepackage{hyperref}
\hypersetup{colorlinks=true,linkcolor=black,citecolor=black,urlcolor=blue,filecolor=black}
\hypersetup{pdfpagemode=UseNone,pdfstartview=}

\usepackage[tableposition=top]{caption}
\usepackage{enumitem}
\setlist[itemize]{noitemsep, topsep=0pt}

\usepackage{apacite}

\advance\oddsidemargin by -0.25in
\advance\textwidth by 0.5in

\advance\topmargin by -0.25in
\advance\textheight by 0.5in

\def\z{\phantom{0}}

\DeclareMathOperator{\thth}{th}

\DeclareMathOperator{\zzNN}{-NN}
\def\kNN#1{{#1}\!\zzNN}
\DeclareMathOperator{\TP}{TP}
\DeclareMathOperator{\FP}{FP}
\DeclareMathOperator{\TN}{TN}
\DeclareMathOperator{\FN}{FN}
\DeclareMathOperator{\TPR}{TPR}
\DeclareMathOperator{\FPR}{FPR}

\long\def\symbolfootnotetext[#1]#2{\begingroup%
\def\thefootnote{\fnsymbol{footnote}}\footnotetext[#1]{#2}\endgroup}

\title{Feature Analysis of Encrypted Malicious Traffic}

\author{Anish Singh Shekhawat\footnotemark[1]\,\,\,\footnotemark[2]\ \ \ 
Fabio Di Troia\footnotemark[1]\,\,\,\footnotemark[3]\ \ \
Mark Stamp\footnotemark[1]\,\,\,\footnotemark[4]}

\begin{document}

\symbolfootnotetext[1]{Department of Computer Science, San Jose State University, San Jose, California, USA}
\symbolfootnotetext[2]{anish.s.shekhawat$@$gmail.com}
\symbolfootnotetext[3]{fabio.ditroia$@$sjsu.edu}
\symbolfootnotetext[4]{mark.stamp$@$sjsu.edu}

\maketitle

\abstract
In recent years there has been a dramatic increase in the number of malware attacks that 
use encrypted HTTP traffic for self-propagation or communication. 
Antivirus software and firewalls typically will 
not have access to encryption keys, and therefore direct detection of
malicious encrypted data is unlikely to succeed. However, previous work has
shown that traffic analysis
can provide indications of malicious intent, even in cases where the underlying data
remains encrypted. In this paper, we apply three machine learning techniques to the problem
of distinguishing malicious encrypted HTTP traffic from benign encrypted traffic
and obtain results comparable to previous work. 
We then consider the problem of feature analysis in some detail.
Previous work has often relied on human expertise to determine the 
most useful and informative features in this problem domain. We demonstrate
that such feature-related information can be obtained directly from machine learning models
themselves. We argue that such a machine learning based approach to feature
analysis is preferable, as it is more reliable, and we can, for example, uncover relatively
unintuitive interactions between features.

\section{Introduction\label{chap:intro}}

Malicious software, or malware, can be defined as a program that is
designed to damage a computer system~\cite{Aycock06}. 
Malware is, arguably, the greatest threat to information security today. 

It is estimated that more than~90\%\ of small-to-medium sized businesses 
have recently experienced an increase in the number of malware detected, 
with some experiencing an increase of~500\%\  in March~2017 alone~\cite{Malwarebytes17}. 
Real-time malware detection based on network traffic has the potential to greatly reduce 
malware propagation on the network.

One approach to detect network-based malware is to use deep packet inspection (DPI).
In DPI, packets are aggregated and the content analyzed to check for signatures or other characteristics
that can be used to classify the data as malicious or benign~\cite{SenSW04}.
Unfortunately, due to the widespread use of the HyperText Transfer Protocol Secure (HTTPS), 
or HTTP over Secure Socket Layer (SSL), straightforward
deep packet inspection methods can be inadequate 
to classify network traffic. It is estimated that HTTPS is used in
more than 70\%\ of Internet traffic today~\cite{Google17}. 

Since HTTPS traffic is encrypted, it cannot be analyzed to the same degree as plaintext traffic. 
This is a benefit if the analyzer is a potential eavesdropper or attacker, but it is harmful
when firewalls are unable to analyze the traffic,
since malware can leverage encryption to evade detection. 
According to a recent report~\cite{Anderson16}, there is a steady~10\%\ to~12\%\ annual increase 
in encrypted malicious network traffic over HTTPS. The~2017 Global Application \&\ Network 
Security Report~\cite{Radware17-18} states that~35\%\ of the organizations
surveyed faced TLS or SSL based attacks, which represents an increase of 50\% over the previous year.

The purpose of this research is to analyze various features that are commonly 
used to distinguish encrypted malicious network traffic from encrypted benign traffic, 
in cases where the decryption keys are unavailable. Specifically, 
we employ machine learning to analyze encrypted network traffic features. 
Again, the emphasis here is on feature analysis, based on trained 
machine learning models. We find that feature analysis based on machine learning models
can provide at least as much useful information---with respect to individual 
features---as human experts can provide. We also show that automated feature 
analysis can easily uncover less intuitive aspects of features. 

The remainder of this paper is organized as follows. Section~\ref{chap:related} gives an overview 
of selected previous work related to the problem of detecting malicious traffic, with the emphasis on 
encrypted traffic. In Section~\ref{chap:dataset}, we introduce the datasets used in our experiments, 
while Section~\ref{chap:method} covers our proposed methodology. 
Section~\ref{chap:experiments} gives our experimental results and analysis,
where the focus is on feature analysis. Finally, in Section~\ref{chap:conclusion}, 
we conclude the paper and outline some areas for future work.

\section{Related Work\label{chap:related}}

Malicious network communication detection typically
relies on either port-based classification or deep packet inspection 
and signature matching. Port-based methods inspect Transmission Control Protocol (TCP) 
and User Datagram Protocol (UDP) port numbers under
the assumption that applications use well-known port numbers~\cite{YoonPPOK09}, 
which are assigned by the Internet Assigned Numbers Authority (IANA)~\cite{IANA}. 
Not surprisingly, malicious applications frequently use non-standard ports in an 
effort to evade such network intrusion detection systems (NIDS) and firewalls 
that rely on similar information~\cite{DregerF06}. 
Even legitimate applications such as Skype use dynamic port numbers to overcome
restrictive firewall policies~\cite{BasetS06}.  In~\cite{MadhukarW06} it is shown that port-based 
classification misclassifies network flow traffic at an astonishingly high rate, estimated to be at 
least~30\%, and possibly as high as~70\%. 

The authors of~\cite{Etienne} use deep packet inspection to detect malicious traffic by
considering payload content and using traditional pattern matching or signature based 
techniques. This research relies on Snort~\cite{Snort}, 
an IDS, to detect malicious traffic, and uses signature or string matching on the packet contents. 
Snort also hosts a popular Intrusion Protection System (IPS) rule set. However, only 
a minuscule percentage of the rules in Snort are TLS-specific, which indicates that such pattern 
matching techniques are not likely to be effective for TLS based malware~\cite{SNORT_TLS}. 

The paper~\cite{SenSW04} demonstrates the use of deep packet inspection to reduce 
false positive and false negative rates when classifying peer-to-peer (P2P) traffic. 
In~\cite{MooreP05}, the authors achieve~100\%\ accuracy when identifying 
network applications, based on an analysis of the entire packet payload. A major
limitation of such methods is the overhead of decrypting and analyzing each packet. 
Of course, these techniques are of not applicable if the 
decryption key is not available to the IDS or IPS.

BotFinder~\cite{TegelerFVK12} is a network flow information analyzer 
that is used to detect bot infections. The system relies on chronologically-ordered flows (or traces) 
to find irregularities in the network behavior between two endpoints. This information, 
along with other network metadata, is used as features in a clustering based 
algorithm~\cite{WangQZ07}. In~\cite{PrasseMPHS17} a neural network based malware detection 
approach is developed, which relies on various network flow features.
The authors of~\cite{LokocKCSP16} present a $k$-nearest neighbor ($k$-NN) 
based classification strategy that is claimed to accurately identify malware
utilizing HTTPS traffic, at least in some specific cases. 

The problem of detecting encrypted Skype traffic is considered in the paper~\cite{MauSar}.
In this work, detection is based on a majority vote of three classifiers, namely, 
a decision tree-based classifier, a logistic classifier, and a Bayesian network classifier.
The system is practical and it appears that relatively strong detection results are obtained. 
Although Skype traffic is not malware, some of the issues that arise when trying to
distinguish encrypted Skype traffic are relevant when attempting to detect 
encrypted malicious traffic.

In~\cite{AndersonM17}, a technique is proposed to identify encrypted malware traffic
based on network flow metadata, using supervised machine learning. These authors 
rely on a complex demilitarized zone (DMZ) architecture to collect the necessary data
for training their machine learning algorithms, 
The traffic collected may not be representative of general network traffic, as this data represents 
only enterprise users. Interestingly,~\cite{AndersonM17} rely heavily on human 
expertise to determine the most important features, which we view as
a significant limitation. Human expertise is not always available, humans are
fallible, and it is possible that non-intuitive features (or non-obvious combinations 
of features) can be as significant to machine learning models---if not more so---than
a collection of seemingly intuitive features.

This paper further explores the use of network flow information as a basis for 
machine learning models that are designed to detect attacks that rely on 
encrypted traffic. The emphasis here is squarely on machine learning---to not only detect and 
distinguish encrypted attack traffic, but to analyze the information that is available 
from the trained models. We show that we can obtain detection results that are 
comparable to previous work. We then consider feature analysis in some depth, 
where our feature analysis is based on trained machine learning models. 
This is in contrast to previous work, which tends to rely primarily 
on domain-specific expertise provided by humans for feature analysis. We believe that
there is significant value in a completely domain-free approach, such as we consider here, 
where essentially no human intervention or domain-specific expertise is required. 
A fully automated system that can rapidly adapt to new and changing circumstances 
would almost certainly require such a domain-free approach. 
In fact, we find that the most informative features differ depending
on the machine learning model used, and some of these features
are not among the most intuitive. Feature interactions are
also difficult to account for when relying on human expertise. 

\section{Dataset}\label{chap:dataset}

In this research, we have used the following two previously published network capture dataset, 
each of which contains both malicious and benign traffic.

\begin{itemize}
\item The CTU-13 dataset~\cite{GarciaGSZ14} was captured as part of research project at
the Czech Technical University. This dataset includes~13 malware traffic captures, 
consisting of both benign and malware traffic. The malware traffic was captured by 
executing selected malware in a Windows virtual machine and recording the 
network traffic generated on the host. The benign traffic was captured on benign 
hosts, i.e., hosts that were not infected with malware. These network captures are 
available as {\tt pcap} files. 
\item
The Malware Capture Facility Project dataset is from another research project carried out 
by the Czech Technical University ATG Group to capture, analyze, and publish 
real malware network traffic~\cite{Erquiaga15}. In this case, the malware was executed with two restrictions,
namely, a bandwidth limit and spam interception. An interesting characteristic of this dataset is that
the malware was allowed to execute over a long period of time---up to several months in some cases. 
Again, the traffic is stored in {\tt pcap} files, and is labeled for ease of use.
\end{itemize}

Our combined dataset includes a total of~72 captures, of which~59 represent malware 
while the remaining~13 are benign. Tables~\ref{tab:con1} and~\ref{tab:flow1} give basic 
statistics for the connections and flows, respectively, in this combined dataset.

\begin{table}[!htb]
	\caption{Dataset connections}\label{tab:con1}
	\centering
		\begin{tabular}{lr}\hline\hline
		\ \ \ \ \ \ \ \ Feature & Count\ \\ \hline
			Benign connection 4-tuples & 8828 \\
			Malicious connection 4-tuples & 52898 \\ \hline
			Total connection 4-tuples & 61726 \\ \hline\hline
		\end{tabular}
\end{table}

\begin{table}[!htb]
	\caption{Dataset flows\label{tab:flow1}}
	\centering
		\begin{tabular}{lr}\hline\hline
		\ \ \ \ \ \ \ \ Feature & Count\ \\ \hline
			Benign flows & 69358 \\
			Malicious flows & 1067273 \\ \hline
			Total flows & 1136631 \\
			\hline\hline
		\end{tabular}
\end{table}


Each capture is contained in a {\tt pcap} file, and includes a list of infected and benign hosts,
as well as Bro IDS logs that were generated from the {\tt pcap} files. Bro~\cite{Bro} is a powerful 
open-source network analysis tool that supports various features 
for traffic inspection, log recording, and attack detection. We use Bro to 
generate network traffic logs that include network flow information and other 
metadata. This information is then used to extract various features related to traffic flows. 
We use the resulting features to train and test our machine learning models. 

Bro generates several types of log files. In this research, we only need to make use of the 
following Bro log files.
\begin{description}
	\item[conn.log] --- This log file contains information about TCP, UDP, and ICMP connections.
	\item[ssl.log] --- The ssl.log file contains information related to SSL/TLS certificates and sessions.
	\item[x509.log] --- As the name suggests, this log file contains relevant information about X.509 
	certificates.
\end{description}

\subsection{Feature Extraction}

\begin{figure}[tb]
	\centering
	\begin{tabular}{c}
		\includegraphics[width=1.0\textwidth]{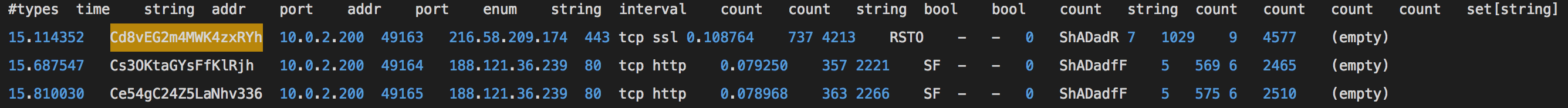} \\
		(a) conn.log \\[2ex]
		\includegraphics[width=1.0\textwidth]{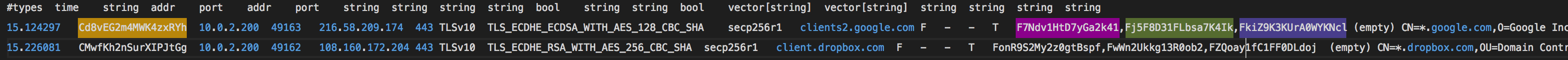} \\
		(b) ssl.log \\[2ex]
		\includegraphics[width=1.0\textwidth]{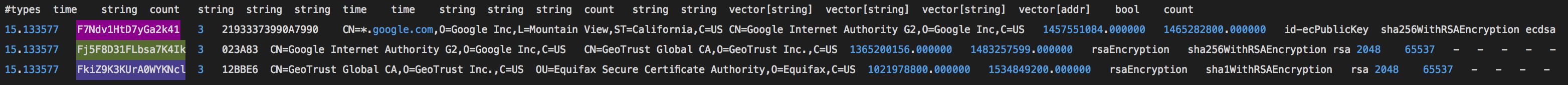} \\
		(c) x509.log
	\end{tabular}
	\vglue-0.05in
	\caption{Interconnection of log records using unique keys in Bro}\label{fig:interconnection}
\end{figure}

We extract several features from Bro logs generated from the network captures under 
consideration. Note that features related to a single connection are generally spread 
over different log files. For example, if there is an SSL connection to a specific server, 
the connection features (e.g., source and destination IP, ports, protocols, duration) 
are stored in the connection log (conn.log). On the other hand, many SSL-specific 
features (e.g., cipher used, server name) are stored in the SSL log (ssl.log), while 
certificate features (e.g., key lengths, common names) are stored in the certificate log (x509.log). 
Bro enables us to deal effectively with the interconnections between the various logs, 
as illustrated in Figure~\ref{fig:interconnection}. Thus, it is relatively easy to extract the 
features that correspond to a given flow.

Bro tracks every incoming and outgoing connection in conn.log, and hence each record in this
log file provides information about a specific connection. Since we are interested only in encrypted 
network connections, we only consider connection records that are related to HTTPS. 
And, since HTTPS connections use SSL/TLS to establish an encrypted link between 
the client and the server, we only extract connection records that have corresponding 
entries in the ssl.log file. 

Every SSL/TLS connection requires a server certificate~\cite{RFC1601}, 
and hence every record in ssl.log contains at least one unique certificate ID 
that the server uses to validate its signing chain. Unique certificate IDs are used to represent 
the certificate record in the x509.log file. We extract only the first certificate ID from the ssl.log file, 
since it corresponds to the end-user certificate, while the remaining IDs correspond to intermediate 
and root certificates.

Every connection record can be identified by a 4-tuple of the form
$$
  (\mbox{source IP}, \mbox{destination IP}, \mbox{destination port}, \mbox{protocol}) . 
$$
We use these 4-tuples as keys to group network features by connection. 

\subsection{Features}

We use Bro to extract a wide variety of features from the connection, SSL, and certificate log files.
At this point, we want as many features as we can reasonable collect. Subsequently, we perform
feature analysis to determine the relative importance of these features,
as well as considering interactions between features.

Table~\ref{tab:2} lists the specific features we extract from the conn.log, the ssl.log, 
and x509.log files. Note that in this table, we use~$\mu$ to represent the mean 
and~$\sigma$ for the standard deviation. Again, all features are determined over a 
single connection, and aggregated based on a specific 4-tuple, as discussed above.

\begin{table}[!htb]
	\caption{Extracted features from conn.log, ssl.log, and x509.log}\label{tab:2}
	\centering
{\scriptsize
		\begin{tabular}{clll}\hline\hline
			  Num & Feature Name & Log & Description \\ \hline
			\z1 & {\tt no\_of\_flows} & conn & Number of records in 4-tuples\\
			\z2 & {\tt avg\_of\_duration} & conn & $\mu$ duration of connections \\
			\z3 & {\tt standard\_deviation\_duration} & conn & $\sigma$ of connections\\
			\z4 & {\tt percent\_sd\_of\_duration} & conn & Percent exceeding $\sigma$\\
			\z5 & {\tt size\_of\_orig\_flows} & conn & Bytes sent by the originator\\
			\z6 & {\tt size\_of\_resp\_flows} & conn & Bytes sent by the responder\\
			\z7 & {\tt ratio\_of\_sizes} & conn & Ratio of responder bytes \\
			\z8 & {\tt percent\_of\_established\_states} & conn & Percent established connections\\
			\z9 & {\tt inbound\_pckts} & conn & Number of incoming packets\\
			10 & {\tt outbound\_pckts} & conn & Number of outgoing packets\\
			11 & {\tt periodicity\_average} & conn & $\mu$ periodicity of connection\\
			12 & {\tt periodicity\_standard\_deviation} & conn & $\sigma$ of connection periodicity\\
			13 & {\tt ssl\_ratio} & ssl & Ratio SSL to non-SSL records\\
			14 & {\tt avg\_key\_len} & ssl & Average key length\\
			15 & {\tt tls\_version\_ratio} & ssl & Ratio of records with TLS \\
			16 & {\tt avg\_of\_certificate\_len} & x509 & Average length\\
			17 & {\tt standart\_deviation\_cert\_len} & x509 & $\sigma$ certificate length\\
			18 & {\tt is\_valid\_certificate} & x509 & 1 if certificate is valid\\
			19 & {\tt amount\_diff\_certificates} & x509 & Number of certificates\\
			20 & {\tt no\_of\_domains\_in\_cert} & x509 & Number of domains\\
			21 & {\tt no\_of\_cert\_path} & x509 & Number of signed paths\\
			22 & {\tt x509\_ssl\_ratio} & x509 & Ratio of SSL logs with x509\\
			23 & {\tt SNI\_ssl\_ratio} & ssl & Ratio with SNI in SSL record\\
			24 & {\tt self\_signed\_ratio} & ssl & Ration with self-signed certificate\\
			25 & {\tt is\_SNIs\_in\_SAN\_dns} & x509 & Check if SNI is SAN DNS\\
			26 & {\tt is\_CNs\_in\_SAN\_dns} & x509 & 1 if all common names in SAN\\
			27 & {\tt differ\_SNI\_in\_ssl\_log} & ssl & Ratio SSL with different SNI\\
			28 & {\tt differ\_subject\_in\_ssl\_log} & ssl & Ratio SSL with different subject\\
			29 & {\tt differ\_issuer\_in\_ssl\_log} & ssl & Ratio SSL with different issuer\\
			30 & {\tt differ\_subject\_in\_cert} & x509 & Ratio subjects \\
			31 & {\tt differ\_issuer\_in\_cert} & x509 & Ratio issuers \\
			32 & {\tt differ\_sandns\_in\_cert} & x509 & Ratio SAN DNS \\
			33 & {\tt ratio\_of\_same\_subjects} & ssl & Ratio SSL with same subject\\
			34 & {\tt ratio\_of\_same\_issuer} & ssl & Ratio SSL with same issuer\\
			35 & {\tt is\_same\_CN\_and\_SNI} & x509 & Checks if CN is same as SNI\\
			36 & {\tt average\_certificate\_expo} &  x509 &Average exponent\\
			37 & {\tt is\_SNI\_top\_level\_domain} & ssl & 1 if SNI is a top level domain\\
			38 & {\tt ratio\_certificate\_path\_error} & x509 & Check if path is valid\\
			\hline\hline
		\end{tabular}
}
\end{table}

\subsection{Labels}

The datasets from~\cite{GarciaGSZ14} and~\cite{Erquiaga15} contains IP addresses of infected and 
benign hosts. We use these IP addresses to label our dataset. That is, if a connection record has an 
infected source IP address, then the record is labeled as malware; otherwise it is labeled as benign.
Of course, these labels are used in both the training phase and in the testing phase.

\section{Methodolgy\label{chap:method}}

We experiment with three machine learning algorithms.
Specifically, we consider experiments involving
support vector machines (SVM), random forests (RF), and
extreme gradient boosting (XGBoost). 
Next, we briefly discuss each of these machine learning techniques,
followed by an introduction to the evaluation metrics that we 
use to quantify our experiments results.

%
%

\subsection{Support Vector Machines}


According to~\cite{BenCamp}, 
support vector machines ``are a rare example of a methodology where geometric
intuition, elegant mathematics, theoretical guarantees, and
practical algorithms meet.'' In particular, the geometric intuition that underlies the topic
is especially helpful for understanding the basic concepts behind SVMs.

For binary classification, 
the key ideas behind SVMs are the following.
\begin{itemize}
\item Separating hyperplane --- The goal is to separate the labeled training data 
into two classes based on a hyperplane.
\item Maximize the margin --- When constructing the separating hyperplane, 
we maximize the ``margin,''
that is, we maximize the minimum separation between the two classes.
\item Work in a higher dimensional space --- We
often try to reduce the dimensionality of data, due to the so-called 
curse of dimensionality. However, when training an SVM, 
it is usually beneficial to work in a higher dimensional space.
By moving the problem to a higher dimension, we have more space
available, and hence there is a better chance of finding
a separating hyperplane.
\item Kernel trick --- In an SVM, we use a kernel function
to transform the data, typically, to a higher dimensional space, with the goal of 
obtaining better separation. The ``trick'' lies in the fact that we pay essentially
no performance penalty for working in this higher dimensional space.
\end{itemize}

By choosing a separating hyperplane that maximizes the margin, 
we give ourselves the largest possible margin for error,
with respect to the training data. This assumes that errors 
are equally likely (and equal in magnitude) in either direction.

SVM training relies on two techniques to deal with training data 
that is not linearly separable. A so-called ``soft'' margin
allows for some classification errors when determining
a separating hyperplane. The more classification errors we
can tolerate, in general, the larger the margin. 
When training an SVMs, there is a user-defined parameter 
that specifies the allowable ``softness'' of the margin.

Another technique employed in SVM training  
is to map the input data to a feature space,
where the problem of constructing a separating hyperplane is more tractable. 
As previously mentioned, this generally involves transforming the input
data to a feature space of higher dimension. 
In a higher dimensional space, it is more likely that 
a separating hyperplane can be found.



For more information on SVMs, see, for example,~\cite{CT,StampML2017}.
In particular,~\cite{BenCamp} is highly recommended.

\subsection{Random Forest}\label{sect:RF}

A random forest can be viewed as a generalization of a decision tree.
To illustrate a decision tree, suppose
that we have a labeled training set consisting of
malware samples and benign samples. From this
training set, we observe that malware samples tend to be smaller in size
and have higher entropy, as compared to benign samples.
We could use this information to construct the decision tree in 
Figure~\ref{fig:mini_DT_mal}, where the thresholds 
for ``large'' versus ``small'' (size) and ``high'' versus ``low'' (entropy) 
would be based on the training data. This decision tree could then 
be used to classify any sample as either malware or benign, 
based on its size and entropy.

\input figures/figure_s4_1.tex

We might want to consider the features in a different order. 
For example, the two features of file size and entropy, as illustrated
in Figure~\ref{fig:mini_DT_mal}, could instead be considered
in the opposite order, as illustrated in Figure~\ref{fig:mini_DT_mal_opp}.

\input figures/figure_s4_2.tex

In general, splits made closer to the root of the tree will tend to 
have more impact on the final classification. 
Therefore, we want to make decisions that are based on the most distinguishing
features closer to the root of the decision tree. In this way,
the decisions for which the training data is less useful are made
later in the process, where such decisions will have less influence on the
final classification.

Information gain is defined as the expected
reduction in entropy when we branch on a given feature.
In the context of a decision tree, information gain
can be computed as the entropy of the parent node minus the average 
weighted entropy of its child nodes.
We can  measure the information gain for each feature, 
and select features in a greedy manner.
In this way, features with the highest gain will be closest
to the root. This is desirable, since the resulting tree will reduce the entropy
as rapidly as possible, which enables us to simplify the tree
by trimming features that provide little or no gain.

Among the advantages of decision trees are simplicity and clarity,
where ``clarity'' means that the tree itself is informative---such
is not always the case for machine learning models.
Of course, there are also some disadvantages to decisions trees,
chief of which is a tendency to overfit the training data. 
That is, the simple and intuitive nature of the tree structure
captures the information in the training data too well, in the sense
that it does not generalize. In machine learning,
this is undesirable, as the training
data is only a representative sample, and we want our models 
to capture the significant properties of this data.

One way to improve on a decision tree is to train multiple trees.
We can select different (overlapping) subsets of the 
training data and construct a tree for each subset, then use a majority vote of the
resulting trees to determine the classification. 
This process is known as bagging
the observations, and it is less likely to overfit,
since it tends to better generalize the training data.

In a random forest, the idea of
bagging is taken one step further---in addition to bagging the observations, 
we also bag the features. That is, we construct multiple decision trees from
selections (with replacement) of the training data,
and also for various selections of the features
(i.e., various subsets and orderings of the features). 
Then for classification, we combine the output from all of 
the resulting decision trees to obtain the final classification. 
For example, we could generate~$t$ decision 
trees, where~$t$ is odd, with each based on a different subset of features 
and data. Then we could use a simple majority vote of the resulting 
trees for classification.

As previously mentioned, a significant advantage of 
a random forest is that it is not prone to overfitting, as compared to a simple decision tree. 
However, random forests do lose some of the inherent simplicity 
and intuitiveness of decision trees.

We note in passing that there is a deep connection between 
$k$-nearest neighbors ($\kNN{k}$)
and random forests. Both~$\kNN{k}$
and decision trees are neighborhood-based 
classification algorithms, but with different neighborhood structures. Since a random forest 
is a collection of decision trees, a random forest is also a neighborhood-based
classification technique, but with a fairly complex
neighborhood structure~\cite{StampML2017}.

For the basics on random forests, a good source is~\cite{BC},
while~\cite{LW} also provides a gentle and practical introduction to the topic. 
For an in-depth discussion of
the connection between random forests and~$\kNN{k}$ 
(and related algorithms), see~\cite{LJ} or~\cite{BC}.

\subsection{XGBoost}\label{sect:boostXG}

Boosting is a process whereby multiple (weak) classifiers are combined
into one, much stronger classifier~\cite{StampML2017}. 
Of course, many machine learning
techniques can be applied in a somewhat analogous fashion. 
For example, SVMs are frequently used to construct a classifier
based on a collection of other machine learning based scores~\cite{SinghTVAS16}. 
The advantage of boosting is
that the individual classifiers can be extremely weak---anything that is better 
than a coin flip can be used. And, provided that we have a sufficient number of 
usable classifiers, boosting enables us to construct an 
arbitrarily strong classifier. 

The best-known boosting algorithm is AdaBoost
which is shorthand for ``adaptive boosting.'' At each iteration 
of AdaBoost, we use a greedy strategy,
in the sense that we select the individual classifier, and its 
associated weight, that improves our overall classifier the most. 
AdaBoost is an adaptive approach, since we build each intermediate 
classifier based on the classifier that was determined 
at the previous step of the algorithm.
It is worth noting that AdaBoost is not a hill climb---we select the 
best available classifier at each step in a greedy manner, but there is
no guarantee that this selection will improve our overall
classifier. 

The results of a simple boosting experiments are given in~\cite{boost}.
In this particular experiment, 1000 classifiers are available, each of which is extremely weak,
with each classifier only providing the correct classification~51\%\ to 52\%\ of the time.
The results of the boosting experiments in~\cite{boost}
are reproduced here in Figure~\ref{fig:ccIter1001000}, where~$L$ is the
number of classifiers used.
The red line illustrates the case where~$L=1000$ classifiers are used,
that is, all of the available classifiers are used. 
The blue line represents the case where~$L=500$ of the~1000
available classifiers are used, and the green line is the case where only~$L=250$ 
of the~1000 classifiers are used.  In each case, the classification accuracy of the 
intermediate (boosted) classifiers are graphed for~$200$ iterations of AdaBoost.

\input figures/fig1001000.tex

From Figure~\ref{fig:ccIter1001000}, we see that with~$L=1000$ classifiers,
we can obtain ideal accuracy using only about~100 of the available classifiers. 
On the other hand, with~$L=500$ classifiers available, we need about~140
iterations before we achieve ideal classification, and with only~$L=250$ 
weak classifiers, we never achieve more than about~90\%\ accuracy.
This illustrates that may need a large number of (weak) classifiers to
achieve impressive results from a boosting algorithm.

Gradient boosting relies on a gradient descent approach to
speed convergence and thereby improve performance.
Extreme gradient boosting, or XGBoost, is a 
highly optimized version of gradient boosting~\cite{ChenG16}.
While AdaBoost is well-known and relatively simple, 
XGBoost has some potential advantage, in terms of efficiency, and additional flexibility
in selecting a cost function~\cite{ChenG16}.

%

\subsection{Cross Validation}

Cross validation consists of partitioning the available data into~$n$ 
equal sized subsets and training~$n$ models, where a different subset 
is reserved for testing in each of these~$n$ ``folds.'' Cross validation serves 
to minimize the effect of bias in the training data, while also maximizing the
number of independent tests on the available data. In this paper, 
we use 10-fold cross validation for all experiments.

\subsection{Evaluation Metrics}

We use accuracy as an evaluation metric for our classification experiments. 
For a given experiment on a labeled data set, 
$$
  \mbox{accuracy} = \dfrac{\TP + \TN}{\TP + \TN + \FP + \FN}
$$
where
\begin{align*}
  \TP &= \mbox{true positives} \\
  \TN &= \mbox{true negatives} \\
  \FP &= \mbox{false positives} \\
  \FN &= \mbox{false negatives}
\end{align*}
That is, the accuracy is simply the ratio of correct classifications to the total number of classifications.

We also employ receiver operating characteristic (ROC) curves in our data analysis.
ROC curves originated with radar engineers during World War~II,
but they are now widely used in many fields. Given a binary classifier, we construct an
ROC curve by plotting the~$\TPR$ versus the~$\FPR$ 
as the threshold varies through the range of data values. 
Equivalently, we plot $1 - \mbox{specificity}$ versus sensitivity
as the threshold varies.

For any ROC curve, the area under the curve (AUC)
varies between~0.0 and~1.0. An AUC of~1.0 indicates ideal separation,
i.e., a threshold exists such that no false positives or false negatives occur. On the other hand,
an AUC of~0.5 indicates that the corresponding binary classifier is no better than flipping a coin.
The AUC can be interpreted as the probability that a randomly selected match case scores higher
than a randomly selected nomatch case~\cite{bradley}.
Whenever we have an AUC of~$x < 0.5$, 
we can simply reverse the match and nomatch criteria 
to obtain a classifier with an AUC of~$1.0-x > 0.5$. 
That is, no binary classifier can do worse than flipping a fair coin.





We note in passing that it is sometimes claimed 
that the AUC tends to overstate the success of experiments. 
However, as previously mentioned,
the AUC measures the probability that a randomly selected positive instance scores 
higher than a randomly selected negative instance---nothing more nor less.
An advantage of the AUC is that it allows us to directly
compare different experimental results, without reference to any specific
threshold value.

\section{Experiments\label{chap:experiments}}

In this section, we present our experimental results. 
Our primary focus is on feature analysis, but as a verification of
the soundness of our approach, we also
show that we can obtain detection and classification results
that are comparable to previous work.
First, we train an SVM classifier
and determine the resulting detection accuracy, and we perform feature analysis 
based on linear SVMs. Then we consider the accuracy of a random forest classifier
and again perform feature analysis. As a third test case, we apply XGBoost
to our data and determine the accuracy for this technique,
and perform feature analysis yet again.
As a final experiment, we consider a multiclass problem, where we
attempt to classify malicious samples into their respective families.
We also consider pairs of features and show that there are 
some interesting interactions among the features.

%

\subsection{SVM Experiments}\label{sect:SVM_exp}

We conducted a variety of experiment based on support vector machines (SVM).
For our first set of experiments, we use a linear kernel and achieve a detection accuracy 
slightly above~92\%. This experiment shows that by using a linear SVM, we are
able to successfully separate encrypted malicious network traffic from
encrypted benign traffic with reasonably high accuracy.

%
%

We also experimented with other kernel functions, specifically, 
radial basis function (RBF) and polynomial kernels.
These results---along with the linear kernel result---are presented in 
the form of a bar graph in Figure~\ref{fig:result_svm_kernels}. 
We observe that the results are not substantially different for 
any of the kernels. In the remainder of this section, 
we only consider a linear kernel, as this greatly simplifies 
feature analysis.


\input figures/SVM_kernels.tex

A linear SVM assigns weights to each input feature, where the weight indicate the 
significance that the SVM places on that feature~\cite{StampML2017}. 
In Figure~\ref{fig:result_svm_weights}, we see that the linear SVM 
assigned the highest weight to feature~12, 
where the features are numbered as in Table~\ref{tab:2}. 
Table~\ref{tab:7} provides a listing of the top~15 of the~38 features, 
based on the linear SVM weights.
From this ranking, we see that average certificate length, periodicity, 
and the average public key length are the
most significant features, at least from the perspective of a linear SVM. 
From~\cite{AndersonM16}, we expect that malware uses weaker 
encryption techniques, such as shorter key lengths and that 
the traffic tends to be more periodic than other applications.
Our SVM feature analysis confirms that the SVM does indeed 
find these same types of features to be most significant.
In addition, the validity of certificate is the sixth highest rank feature, 
which tell us that malware often does not use a valid certificate.
This is also consistent with manual analysis of the data.


\input figures/SVM_weights.tex

\begin{table}[!htb]
	\caption{Top~15 features as ranked by linear SVM}\label{tab:7}
	\centering
		\begin{tabular}{cl}\hline\hline
			Rank & \ \ \ Feature\\ \hline
\z1   & {\tt periodicity\_standard\_deviation} \\ 
\z2  & {\tt periodicity\_average} \\ 
\z3  & {\tt avg\_of\_certificate\_length} \\ 
\z4  & {\tt avg\_of\_duration} \\ 
\z5  & {\tt standard\_deviation\_duration} \\ 
\z6  & {\tt is\_valid\_certificate} \\ 
\z7  & {\tt is\_SNIs\_in\_SAN\_dns} \\ 
\z8  & {\tt avg\_key\_len} \\ 
\z9  & {\tt standard\_deviation\_cert\_length} \\ 
10  & {\tt self\_signed\_ratio} \\ 
11 & {\tt amount\_diff\_certificates} \\ 
12 & {\tt x509\_ssl\_ratio} \\ 
13 & {\tt tls\_version\_ratio} \\ 
14 & {\tt ratio\_of\_same\_subjects} \\ 
15 & {\tt percent\_of\_established\_states} \\ 
			 \hline\hline
		\end{tabular}
\end{table}



%


Next, we consider recursive feature elimination (RFE).
In RFE, we iteratively eliminate the lowest ranking feature, 
then train a new model on the reduced feature set. 
In this way, interactions between the features are
accounted for, which is not the case with the feature ranking
in Table~\ref{tab:7}.

From Figure~\ref{fig:result_rfe}, we see that by using just the
top~6 features obtained from an RFE (based on linear SVMs), we can obtain
results that are within~2\%\ of those obtained
using the full set of~38 features. Furthermore, if we use the top~10 features,
our accuracy is within~1\%\ of the the results obtained when 
using all~38 features. These results demonstrate that for this problem,
a small number of features contain most of the discriminatory strength.


\input figures/SVM_RFE.tex

In Table~\ref{tab:7RFE}, we have listed the top~15 of the~38 features, 
as determined by RFE, based on a linear SVM.
Comparing these results to the ranked features in Table~\ref{tab:7},
we see significant differences---for example, the top ranked
feature by the linear weight ranking is only the fifth highest ranked
feature according to the RFE. In addition, the 8th ranked feature
according to the RFE does not appear among the to~15
features, based on a simple feature ranking.
In general, we expect the RFE ranking to be more meaningful, 
as the RFE better accounts for interactions between features.

\begin{table}[!htb]
	\caption{Top~15 features by SVM RFE ranking}\label{tab:7RFE}
	\centering
		\begin{tabular}{cl}\hline\hline
			Rank & \ \ \ Feature \\ \hline
			 \z1  & {\tt periodicity\_standard\_deviation} \\
			 \z2  & {\tt periodicity\_average} \\
			 \z3  & {\tt avg\_of\_duration} \\
			 \z4  & {\tt is\_SNIs\_in\_SAN\_dns} \\
			 \z5  & {\tt avg\_of\_certificate\_length} \\
			 \z6  & {\tt avg\_key\_len} \\
			 \z7  & {\tt self\_signed\_ratio} \\
			 \z8  &  {\tt is\_same\_CN\_and\_SNI} \\
			 \z9  &  {\tt percent\_of\_established\_states} \\
			 10 &  {\tt differ\_issuer\_in\_ssl\_log} \\
                           11 &  {\tt differ\_subject\_in\_cert} \\
                           12 &  {\tt ratio\_certificate\_path\_error} \\
                           13 &  {\tt ratio\_of\_same\_issuer} \\
                           14 &  {\tt size\_of\_resp\_flows} \\ 
                           15 &  {\tt ratio\_of\_sizes} \\
                           \hline\hline
		\end{tabular}
\end{table}

\subsection{Random Forest Experiments}

Next, we consider a random forest classifier. The key parameter for a random
forest is the number of estimators, that is, the number of bagged
decision trees (see Section~\ref{sect:RF} for a discussion of bagging). 
In Figure~\ref{fig:result_rf_estimators}, we have graphed 
the accuracy of the random forest as a function of the number of estimators.
Interestingly, only a relatively small number of estimators are needed to achieve good
accuracy. However, to ensure optimal results, and since there is little additional
training cost associated with a larger number of estimators, 
we use~500 estimators for all random forest experiments presented in this section.


We again performed recursive feature elimination, but based on a
random forest classifier rather than a linear SVM.
The features rankings obtained from this random forest RFE are given in 
Figure~\ref{fig:result_rf_feature_imp}. 



\input figures/RF_estimators.tex

\input figures/RF_feature_importance.tex

We observe that the feature ranking in the case of random forest
differ significantly from the rankings generated using a linear SVM,
as discussed in Section~\ref{sect:SVM_exp}. 
From Figure~\ref{fig:result_rf_feature_imp}, we see that
the top four features from the random forest classifier stand 
out from the remaining features---these four features are listed 
in Table~\ref{tab:RF_RFE}. Interestingly, of these four dominant
random forest features, only the fourth appears in the list of top ranked
SVM features, as given in Table~\ref{tab:7RFE},
and this feature only ranks as the~$15^{\thth}$ best according to 
our SVM-based analysis.

\begin{table}[!htb]
	\caption{Top 4 features from random forest RFE}\label{tab:RF_RFE}
	\centering
		\begin{tabular}{cl}\hline\hline
			Rank & \ \ \ Feature \\ \hline
			\z1 & {\tt size\_of\_orig\_flows} \\
			\z2 & {\tt no\_of\_cert\_path} \\			 
			\z3 & {\tt standard\_deviation\_duration} \\
			\z4 & {\tt ratio\_of\_sizes} \\
                           \hline\hline
		\end{tabular}
\end{table}

Using the top ranked features from the RFE, as listed in Figure~\ref{fig:result_rf_feature_imp},
we obtain the accuracy results in Figure~\ref{fig:result_rf_rfe}.
In this case, we obtain a maximum accuracy of nearly~99\%,
which is significantly better than a linear SVM, and we observe that
we have an accuracy in excess of~98.5\%\ with just~6 features.


\input figures/RF_RFE.tex

We note that the highly ranked random forest features are less
intuitive than the SVM features, yet we can obtain stronger
results with the random forest based on these features, 
as compared to an SVM. This nicely illustrates the importance
of letting the model dictate features selection, rather than relying
on the most intuitively appealing features.

\subsection{XGBoost Experiments}

Finally, we construct classifiers using XGBoost, based on
decision trees. In Figure~\ref{fig:result_xgboost_estimators}, we give our
XGBoost results as a function of the number of estimators (i.e., decision tress). 
In light of the discussion of boosting in Section~\ref{sect:boostXG},
it should not be surprising that a large number of estimators are required
before we obtain near-optimal accuracy. 
In all of the XGBoost experiments discussed in the remainder of this section,
we use~1000 estimators.


\input figures/XG_estimators.tex

We performed recursive feature elimination using XGBoost---these 
results are summarized in Figure~\ref{fig:result_xgboost_feature_imp}.
Note that the top ranked XGBoost features differ somewhat from those obtained with
the random forest, as give in Figure~\ref{fig:result_rf_rfe}. However, the
feature rankings provided by our random forest and SGBoost experiments
are more similar to each other than either is to the SVM rankings.
That is, the SVM feature rankings can be viewed as the outlier
among our three sets of experiments.


\input figures/XG_feature_importance.tex

In Figure~\ref{fig:result_xgboost_rfe}, we give the accuracy of XGBoost as a 
function of the top ranked features. As with a random forest, we see that a small
number of features is sufficient to obtain near-optimal accuracy. 


\input figures/XG_RFE.tex

The results in Figure~\ref{fig:result_xgboost_rfe} serve to
reinforce the point that the highly intuitive features determined
by the SVM classifier are not the optimal set for
constructing a classifier, particularly in case we want to 
minimize the number of features. In practice, we would
want to reduce the number of features that need to be collected,
as this will enable more efficient scoring. Efficiency is particularly 
important for the problem at hand, since we are dealing with
network traffic that would need to be analyzed in real time.


Before briefly turning our attention to the multiclass problem,
we summarize our SVM, random forest, and boosting experiments. 
A comparison of the classification accuracy for these
machine learning algorithms 
is given in the form of a bar graph in Figure~\ref{fig:result_svm_rf_xgb}.
We see that XGBoost gives the highest accuracy, at~99.15\%,
while random forest is virtually indistinguishable at~98.78\%\ accuracy, 
with a linear SVM performing significantly worse.


\input figures/ML_comparison.tex

\subsection{Multiclass Family Classification}

For our multiclass experiments, we attempt to classify samples
from four malware families, namely, Dridex, Trickbot, WannaCry, and Zbot.
The data for these experiments is summarized in Table~\ref{tab:8}.


\begin{table}[!htb]
	\caption{Malware families}\label{tab:8}
	\centering
		\begin{tabular}{c|cc}\hline\hline
			Family & Connection 4-tuples & Flows \\ \hline
			Dridex & \z24 & \z65465 \\
			Trickbot & 217 & 465289 \\
			WannaCry & \z34 & \z\z\z785 \\
			Zbot & \z45 & \z\z1788 \\ \hline\hline
		\end{tabular}
\end{table}

We trained models for all six pairs of the four malware families listed in Table~\ref{tab:8}, 
and we tested each model independently. These results are summarized in 
Figure~\ref{fig:result_multiclass}, where we see that a random forest performs better 
than either an SVM or XGboost in most cases, with XGBoost giving us higher accuracy 
only for the Dridex versus Trickbot case. However, the differences are 
relatively small in every case.


\input figures/classification.tex

We also trained true multiclass models, that is, we comsidered data from all four families 
together for training. The results of these experiments are given in 
Figure~\ref{fig:result_multiclass_all}, where we see that a random forest performs 
best, but only marginally better than XGBoost. Perhaps the most interesting
aspect of this experiment is the overall high accuracy, given that we are classifying
samples from four diverse malware families. These results indicate that although the 
families are significantly different from each other, they are generally more different
from the benign traffic than they are from each other.
This is a significant observation, since it implies that we are likely to be
able to filter malicious encrypted traffic from benign traffic with one model,
instead of needing to rely on a model for each individual family. This
would make detection based on machine learning models 
far more practical than a situation where a multitude of different 
models are required.


\input figures/multiclass.tex

\subsection{Discussion}

The accuracy results in the previous section show that XGBoost performs better than 
the other two algorithms considered, but only marginally better than a random forest. 
In addition to the accuracy numbers cited above, we
have computed the area under the ROC curve (AUC). For the AUC statistic, 
we obtain a value of~$0.9122$ for the linear SVM,
we have~$0.9980$ for a random forest,
while XGBoost achieves an AUC of~$0.9988$.
The corresponding ROC curve for these SVM, random forest, 
and XGBoost experiments are given in Figure~\ref{fig:ROC}~(a), 
(b), and~(c), respectively. These AUC values serve to further reinforce 
the point that the random forest and XGBoost offer virtually the same performance
on our dataset, while also providing a more detailed view of the relative weakness of the SVM. 

\input figures/fig_ROC_curves.tex

In Figure~\ref{fig:ml_comparison}, we give a summary of several 
measures of success for our experiments.
These results confirm that our results are
comparable to---if not better than---those achieved in any of the relevant previous
work cited in Section~\ref{chap:related}. 




\input figures/measures_compared.tex

Feature analysis is the main point of our work, and the experiments above show
that we can determine strong sets of features directly from the machine learning
models. It is interesting to note that the strongest features for our best models 
are not the most intuitive features. This illustrates the strength of a domain-free
approach to feature selection, as opposed to relying on the most intuitively appealing
features.



In an attempt to gain further insight into the various features, 
we briefly consider feature interactions. Figure~\ref{fig:heatmap} 
gives the Pearson correlation coefficient~\cite{Pearson} 
for all pairs of features in the form of a heatmap. Note that the feature numbers
in this heatmap correspond to the feature numbers in Table~\ref{tab:2}.


\input figures/heatmap.tex

From the heatmap in Figure~\ref{fig:heatmap} we see that, 
for example, features~14 through~18 are all highly
correlated. These features include
the length of the encryption key, the certificate length, 
and the validity of the certificate. The fact that these features are 
correlated provides evidence that substandard encryption 
is generally employed by encrypted malware. 
We also note that periodicity
properties, which appear as features~11 and~12,
are uncorrelated with the encryption features, as given in
features~14 through~18. In fact, these periodicity features are
uncorrelated with almost all other features. 

It is also interesting to consider, for example, the four strongest features
of the random forest, as given in Table~\ref{tab:RF_RFE}. These
four features are, in order, numbers~5, 21, 3, and~7. From 
the heatmap in Figure~\ref{fig:heatmap}, we see that these features
are only weakly correlated with each other. This provides further
evidence that the random forest RFE is able to determine a strong set of 
features that are relatively independent of each other, and hence these
features are providing independent information to the random forest.

On the other hand, if we look closely at the more intuitive
features determined by the SVM RFE, we find some strong correlations.
For example, the periodicity features (numbers~11 and~12) are the top two
features in the SVM, yet they are highly correlated, and hence
including both of these features provides very little new information,
as compared to only including either one. This is significant as, again, 
fewer features will enable faster and more efficient scoring.
This also points out a risk of relying on domain-specific knowledge for
feature extraction---correlations are difficult to account for in such an
approach, and hence some level of redundancy is highly likely.



\section{Conclusion and Future Work\label{chap:conclusion}}


With the widespread use of HTTPS and advancement in malware detection techniques, 
there has been a rapid increase in the number of malware samples using HTTPS encryption 
to evade detection---a trend that is sure to continue into the future.
This is worrying because encryption disrupts the most popular and effective 
malware detection techniques available today. 

In this research, we considered the challenging problem of classifying encrypted network traffic 
as malicious or benign, without any decryption or deep packet inspection. Our focus here 
has been squarely on the important problem of feature analysis---a topic that we believe 
has been neglected in previous research in this field.
We considered three machine learning algorithms, namely, SVM, XGBoost, and random forest. 
These algorithms were used to train and test models. We then performed
feature analysis using RFE in each case. Our results show that XGBoost performed slightly
better than a random forest, with both being about~99\%\ accurate, 
while an SVM performed relatively poorly. Our main contribution is that we provided 
a thorough analysis of the various features considered. We found that a small number of
features suffice, and that the optimal set of features are not necessarily the most intuitive.
This is significant, as a minimal set of features is necessary for a practical and 
efficient implementation of any such detection system. We also argued that feature 
selection based on the models themselves---which we refer to as a domain-free 
approach---is preferable to a domain-specific approach that relies on human
expertise to select features.

Security research is often a double-edged sword, and the feature analysis considered
here is no exception. For efficient detection, it is necessary to know which features contribute 
the most, but this same analysis also points the way toward ``new-and-improved'' malware.
That is, malware writers can use the feature analysis presented here to make
their creations more difficult to detect.
This observation points to the need for more research in this field, 
so that detection techniques can stay ahead of the inevitable improvements
in malware development.


For further research, it would be useful to have a larger dataset that we could mine
for more subtle features. 
Another direction for related research would be to field a system based on
the machine learning techniques considered here,
so that the effectiveness and robustness of such an approach could be analyzed 
under real-world and changing conditions. In particular, it would be interesting
to consider the evolution of such a system over time, as human adversaries
improve their attacks and the detection system then adapts to 
these changes in attack strategies.

\bibliographystyle{apacite}

\bibliography{references}

\end{document}

%% file: figures/figure_s4_1.tex
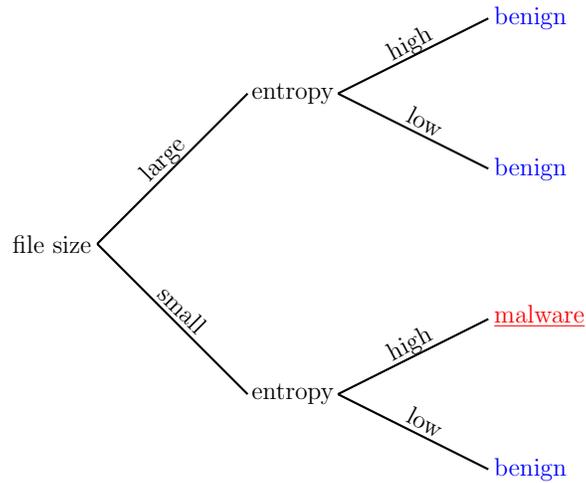
\begin{figure}[!tb]
  \begin{center}
    \begin{tikzpicture}[scale=0.8, every node/.style={scale=0.8}]
    \draw[thick,color=black] (0,0) -- (2.5,2.5);
    \draw[thick,color=black] (0,0) -- (2.5,-2.5);
    \draw[thick,color=black] (4.0,2.5) -- (6.5,3.75);
    \draw[thick,color=black] (4.0,2.5) -- (6.5,1.25);
    \draw[thick,color=black] (4.0,-2.5) -- (6.5,-1.25);
    \draw[thick,color=black] (4.0,-2.5) -- (6.5,-3.75);
    \node at (-0.75,0){file size};
    \node at (3.25,2.5){entropy};
    \node at (3.25,-2.5){entropy};
    \node[color=blue] at (7.2,3.75){benign};
    \node[color=red] at (7.35,-1.2){\underline{malware}};
    \node[color=blue] at (7.2,1.25){benign};
    \node[color=blue] at (7.2,-3.75){benign};

   \node[rotate=45] at (1.1,1.4){large};
   \node[rotate=-45] at (1.415,-1.085){small};
   \node[rotate=28] at (5.2,3.325){high};
   \node[rotate=28] at (5.2,-1.675){high};
   \node[rotate=-28] at (5.45,-2.96){low};
   \node[rotate=-28] at (5.45,2.04){low};

    

    \end{tikzpicture}
  \end{center}
  \vglue-0.2in
  \caption{Decision tree example\label{fig:mini_DT_mal}}
\end{figure}

%% file: figures/figure_s4_2.tex
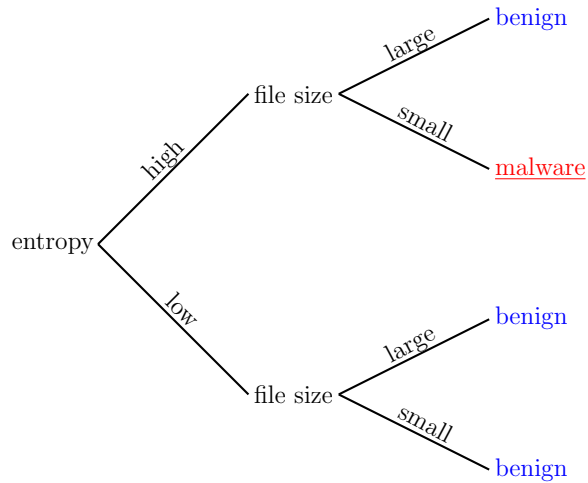
\begin{figure}[!tb]
  \begin{center}
    \begin{tikzpicture}[scale=0.8, every node/.style={scale=0.8}]
    \draw[thick,color=black] (0,0) -- (2.5,2.5);
    \draw[thick,color=black] (0,0) -- (2.5,-2.5);
    \draw[thick,color=black] (4.0,2.5) -- (6.5,3.75);
    \draw[thick,color=black] (4.0,2.5) -- (6.5,1.25);
    \draw[thick,color=black] (4.0,-2.5) -- (6.5,-1.25);
    \draw[thick,color=black] (4.0,-2.5) -- (6.5,-3.75);
    \node at (-0.75,0){entropy};
    \node at (3.25,2.5){file size};
    \node at (3.25,-2.5){file size};
    \node[color=blue] at (7.2,3.75){benign};
    \node[color=red] at (7.35,1.25){\underline{malware}};
    \node[color=blue] at (7.2,-1.225){benign};
    \node[color=blue] at (7.2,-3.75){benign};

   \node[rotate=45] at (1.1,1.4){high};
   \node[rotate=-45] at (1.415,-1.085){low};
   \node[rotate=28] at (5.2,3.325){large};
   \node[rotate=28] at (5.2,-1.675){large};
   \node[rotate=-28] at (5.45,-2.96){small};
   \node[rotate=-28] at (5.45,2.04){small};

    

    \end{tikzpicture}
  \end{center}
  \vglue-0.2in
  \caption{Features in different order\label{fig:mini_DT_mal_opp}}
\end{figure}

%% file: figures/fig1001000.tex
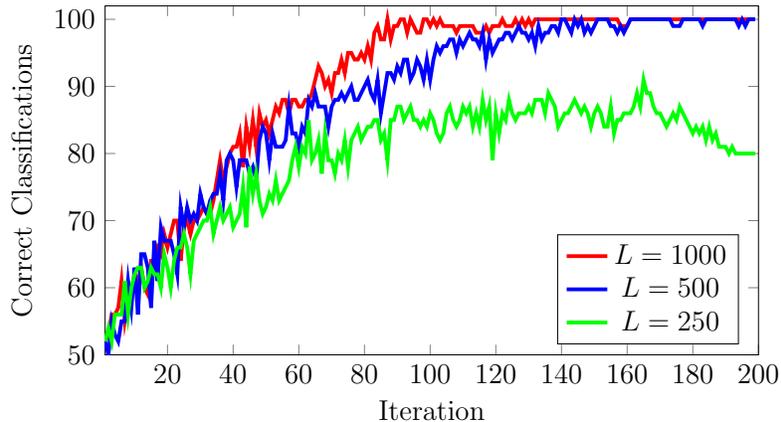
\begin{figure}[!tbp]
\centering
    \begin{tikzpicture}[scale=0.9]
    \begin{axis}[width=0.75\textwidth,height=0.45\textwidth,
    		      xmin=1.0,xmax=200.0,
                       ymin=50.0,ymax=102.0, legend pos=south east,
                       x tick label style={
    			/pgf/number format/.cd,
			/pgf/number format/1000 sep={},
    			fixed,
    			fixed zerofill,
    			precision=0},                       
                       xlabel={Iteration},ylabel={Correct Classifications},
                       y tick label style={
    			/pgf/number format/.cd,
    			fixed,
    			fixed zerofill,
    			precision=0}] 
\addplot[color=red,ultra thick,
	mark=none] coordinates {
(1, 52)
(2, 53)
(3, 56)
(4, 56)
(5, 57)
(6, 61)
(7, 55)
(8, 58)
(9, 59)
(10, 61)
(11, 61)
(12, 63)
(13, 60)
(14, 59)
(15, 64)
(16, 64)
(17, 66)
(18, 65)
(19, 68)
(20, 66)
(21, 68)
(22, 70)
(23, 70)
(24, 64)
(25, 69)
(26, 70)
(27, 68)
(28, 70)
(29, 70)
(30, 71)
(31, 72)
(32, 71)
(33, 71)
(34, 73)
(35, 76)
(36, 79)
(37, 76)
(38, 78)
(39, 80)
(40, 81)
(41, 81)
(42, 83)
(43, 78)
(44, 84)
(45, 81)
(46, 85)
(47, 81)
(48, 86)
(49, 84)
(50, 85)
(51, 83)
(52, 85)
(53, 87)
(54, 86)
(55, 88)
(56, 88)
(57, 86)
(58, 88)
(59, 88)
(60, 87)
(61, 87)
(62, 88)
(63, 88)
(64, 89)
(65, 91)
(66, 93)
(67, 92)
(68, 90)
(69, 91)
(70, 89)
(71, 92)
(72, 92)
(73, 94)
(74, 92)
(75, 95)
(76, 95)
(77, 94)
(78, 94)
(79, 95)
(80, 94)
(81, 96)
(82, 94)
(83, 98)
(84, 99)
(85, 97)
(86, 97)
(87, 100)
(88, 97)
(89, 99)
(90, 99)
(91, 100)
(92, 100)
(93, 98)
(94, 100)
(95, 99)
(96, 98)
(97, 98)
(98, 100)
(99, 100)
(100, 99)
(101, 100)
(102, 99)
(103, 100)
(104, 99)
(105, 99)
(106, 99)
(107, 99)
(108, 98)
(109, 99)
(110, 99)
(111, 99)
(112, 99)
(113, 99)
(114, 98)
(115, 98)
(116, 98)
(117, 99)
(118, 98)
(119, 99)
(120, 99)
(121, 99)
(122, 100)
(123, 99)
(124, 99)
(125, 99)
(126, 100)
(127, 99)
(128, 100)
(129, 99)
(130, 100)
(131, 99)
(132, 100)
(133, 100)
(134, 100)
(135, 100)
(136, 100)
(137, 100)
(138, 100)
(139, 100)
(140, 100)
(141, 100)
(142, 100)
(143, 100)
(144, 100)
(145, 100)
(146, 100)
(147, 100)
(148, 100)
(149, 100)
(150, 100)
(151, 100)
(152, 100)
(153, 100)
(154, 100)
(155, 100)
(156, 100)
(157, 100)
(158, 99)
(159, 99)
(160, 99)
(161, 100)
(162, 100)
(163, 100)
(164, 100)
(165, 100)
(166, 100)
(167, 100)
(168, 100)
(169, 100)
(170, 100)
(171, 100)
(172, 100)
(173, 100)
(174, 100)
(175, 100)
(176, 100)
(177, 100)
(178, 100)
(179, 100)
(180, 100)
(181, 100)
(182, 100)
(183, 100)
(184, 100)
(185, 100)
(186, 100)
(187, 100)
(188, 100)
(189, 100)
(190, 100)
(191, 100)
(192, 100)
(193, 100)
(194, 100)
(195, 100)
(196, 100)
(197, 100)
(198, 100)
(199, 100)
};
\addlegendentry{$L = 1000$}
\addplot[color=blue,ultra thick,
	mark=none] coordinates {
(1, 52)
(2, 50)
(3, 55)
(4, 53)
(5, 52)
(6, 55)
(7, 55)
(8, 61)
(9, 57)
(10, 63)
(11, 56)
(12, 65)
(13, 65)
(14, 63)
(15, 57)
(16, 67)
(17, 61)
(18, 69)
(19, 67)
(20, 67)
(21, 67)
(22, 65)
(23, 62)
(24, 72)
(25, 69)
(26, 72)
(27, 70)
(28, 71)
(29, 70)
(30, 74)
(31, 72)
(32, 71)
(33, 73)
(34, 74)
(35, 74)
(36, 77)
(37, 73)
(38, 79)
(39, 80)
(40, 79)
(41, 75)
(42, 79)
(43, 79)
(44, 79)
(45, 77)
(46, 78)
(47, 76)
(48, 82)
(49, 84)
(50, 83)
(51, 79)
(52, 83)
(53, 81)
(54, 81)
(55, 81)
(56, 82)
(57, 87)
(58, 83)
(59, 83)
(60, 81)
(61, 84)
(62, 83)
(63, 88)
(64, 87)
(65, 89)
(66, 87)
(67, 87)
(68, 87)
(69, 84)
(70, 87)
(71, 88)
(72, 88)
(73, 89)
(74, 88)
(75, 89)
(76, 90)
(77, 88)
(78, 88)
(79, 89)
(80, 90)
(81, 91)
(82, 89)
(83, 88)
(84, 91)
(85, 86)
(86, 89)
(87, 92)
(88, 92)
(89, 91)
(90, 93)
(91, 92)
(92, 94)
(93, 94)
(94, 91)
(95, 92)
(96, 94)
(97, 93)
(98, 91)
(99, 94)
(100, 91)
(101, 94)
(102, 96)
(103, 95)
(104, 96)
(105, 96)
(106, 97)
(107, 97)
(108, 95)
(109, 97)
(110, 96)
(111, 97)
(112, 98)
(113, 97)
(114, 97)
(115, 96)
(116, 98)
(117, 94)
(118, 97)
(119, 95)
(120, 96)
(121, 97)
(122, 97)
(123, 98)
(124, 98)
(125, 97)
(126, 98)
(127, 99)
(128, 98)
(129, 98)
(130, 97)
(131, 98)
(132, 96)
(133, 99)
(134, 97)
(135, 98)
(136, 98)
(137, 99)
(138, 99)
(139, 98)
(140, 100)
(141, 100)
(142, 99)
(143, 98)
(144, 100)
(145, 99)
(146, 100)
(147, 98)
(148, 99)
(149, 98)
(150, 99)
(151, 100)
(152, 100)
(153, 100)
(154, 100)
(155, 99)
(156, 99)
(157, 98)
(158, 99)
(159, 98)
(160, 99)
(161, 100)
(162, 100)
(163, 100)
(164, 100)
(165, 100)
(166, 100)
(167, 100)
(168, 100)
(169, 100)
(170, 100)
(171, 100)
(172, 100)
(173, 100)
(174, 99)
(175, 99)
(176, 99)
(177, 100)
(178, 100)
(179, 100)
(180, 100)
(181, 100)
(182, 100)
(183, 99)
(184, 100)
(185, 100)
(186, 100)
(187, 100)
(188, 100)
(189, 100)
(190, 100)
(191, 100)
(192, 100)
(193, 100)
(194, 99)
(195, 100)
(196, 99)
(197, 100)
(198, 100)
(199, 100)
};
\addlegendentry{$L = 500$}
\addplot[color=green,ultra thick,
	mark=none] coordinates {
(1, 52)
(2, 54)
(3, 52)
(4, 56)
(5, 56)
(6, 56)
(7, 61)
(8, 56)
(9, 60)
(10, 62)
(11, 63)
(12, 63)
(13, 60)
(14, 61)
(15, 63)
(16, 62)
(17, 62)
(18, 60)
(19, 65)
(20, 63)
(21, 60)
(22, 64)
(23, 66)
(24, 66)
(25, 68)
(26, 66)
(27, 62)
(28, 67)
(29, 68)
(30, 69)
(31, 70)
(32, 70)
(33, 72)
(34, 68)
(35, 70)
(36, 72)
(37, 70)
(38, 71)
(39, 72)
(40, 69)
(41, 70)
(42, 71)
(43, 75)
(44, 69)
(45, 78)
(46, 74)
(47, 77)
(48, 73)
(49, 71)
(50, 72)
(51, 74)
(52, 73)
(53, 76)
(54, 73)
(55, 74)
(56, 75)
(57, 76)
(58, 79)
(59, 81)
(60, 78)
(61, 81)
(62, 80)
(63, 85)
(64, 79)
(65, 80)
(66, 82)
(67, 77)
(68, 83)
(69, 79)
(70, 79)
(71, 77)
(72, 80)
(73, 79)
(74, 82)
(75, 78)
(76, 82)
(77, 83)
(78, 84)
(79, 82)
(80, 83)
(81, 84)
(82, 84)
(83, 85)
(84, 85)
(85, 82)
(86, 83)
(87, 79)
(88, 85)
(89, 85)
(90, 87)
(91, 87)
(92, 85)
(93, 86)
(94, 87)
(95, 85)
(96, 85)
(97, 84)
(98, 85)
(99, 84)
(100, 86)
(101, 85)
(102, 83)
(103, 82)
(104, 83)
(105, 86)
(106, 87)
(107, 85)
(108, 84)
(109, 85)
(110, 87)
(111, 86)
(112, 87)
(113, 84)
(114, 86)
(115, 84)
(116, 88)
(117, 85)
(118, 87)
(119, 79)
(120, 86)
(121, 83)
(122, 86)
(123, 85)
(124, 88)
(125, 86)
(126, 85)
(127, 86)
(128, 86)
(129, 86)
(130, 87)
(131, 86)
(132, 88)
(133, 88)
(134, 88)
(135, 86)
(136, 89)
(137, 88)
(138, 87)
(139, 84)
(140, 85)
(141, 86)
(142, 86)
(143, 87)
(144, 88)
(145, 87)
(146, 86)
(147, 87)
(148, 86)
(149, 84)
(150, 86)
(151, 87)
(152, 86)
(153, 84)
(154, 84)
(155, 87)
(156, 83)
(157, 85)
(158, 85)
(159, 86)
(160, 86)
(161, 86)
(162, 87)
(163, 89)
(164, 87)
(165, 91)
(166, 89)
(167, 89)
(168, 87)
(169, 86)
(170, 86)
(171, 85)
(172, 84)
(173, 85)
(174, 86)
(175, 85)
(176, 87)
(177, 86)
(178, 85)
(179, 84)
(180, 83)
(181, 84)
(182, 83)
(183, 84)
(184, 82)
(185, 84)
(186, 82)
(187, 83)
(188, 81)
(189, 81)
(190, 81)
(191, 80)
(192, 81)
(193, 80)
(194, 80)
(195, 80)
(196, 80)
(197, 80)
(198, 80)
(199, 80)
};
\addlegendentry{$L = 250$}
    \end{axis}
    \end{tikzpicture}
\vglue-0.05in
\caption{Correct classifications vs iteration ($n=100$)}\label{fig:ccIter1001000}	
\end{figure}

%% file: figures/SVM_kernels.tex
\begin{figure}[!tbp]
\centering
\begin{tikzpicture}[scale=0.9]
    \begin{axis}[
        width  = 0.625*\textwidth,
        height = 0.5*\textwidth,
        major x tick style = transparent,
        ybar=4*\pgflinewidth,
        bar width=25pt,
        ymajorgrids = true,
        ylabel = {Accuracy},
        symbolic x coords={Linear,RBF,Polynomial},
	y tick label style={
    	/pgf/number format/.cd,
   	fixed,
   	fixed zerofill,
    	precision=2},
        xtick = data,
        x tick label style={rotate=45,anchor=north east, inner sep=0mm},
        scaled y ticks = false,
        enlarge x limits=0.25,
        ymin=0.84,
        ymax=0.94,
        legend cell align=left,
        legend pos=south east,
    ]
        \addplot[fill=blue]
            coordinates {
(Linear,0.920604119)
(RBF,0.926919199)
(Polynomial,0.92149175)
};
    \end{axis}
\end{tikzpicture}
\vglue-0.05in
\caption{SVM with different kernels}\label{fig:result_svm_kernels}
\end{figure}
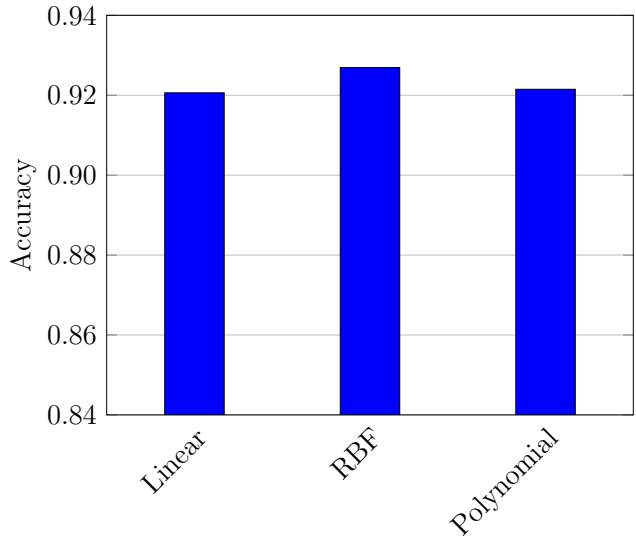

%% file: figures/SVM_weights.tex
\begin{figure}[!tbp]
\centering
\begin{tikzpicture}
    \begin{axis}[
        width  = 0.9*\textwidth,
        height = 7.5cm,
        major x tick style = transparent,
        ybar=4*\pgflinewidth,
        bar width=5pt,
        ymajorgrids = true,
        ylabel = {Accuracy},
        symbolic x coords={1,2,3,4,5,6,7,8,9,10,11,12,13,14,15,16,17,18,19,20,21,22,23,24,25,26,27,28,29,30,31,32,33,34,35,36,37,38},
	y tick label style={
    	/pgf/number format/.cd,
   	fixed,
   	fixed zerofill,
    	precision=1},
        xtick = data,
        x tick label style={inner sep=0mm,font=\tiny},
        scaled y ticks = false,
        enlarge x limits=0.02,
        ymin=0,
        ymax=1.0,
        legend cell align=left,
        legend pos=south east,
    ]
        \addplot[fill=blue]
            coordinates {
(1,0.00750572366935)
(2,0.515466180086)
(3,0.283897786869)
(4,0.00142651070207)
(5,0.00496043958281)
(6,0.000267270536395)
(7,0.00173821582285)
(8,0.0115136018291)
(9,0.0000215117517657)
(10,0.00404953979928)
(11,0.827313774675)
(12,1)
(13,0.00415261614357)
(14,0.136254396962)
(15,0.0131122573555)
(16,0.653392193716)
(17,0.128326527652)
(18,0.208823486547)
(19,0.0160123874839)
(20,0.0100198305461)
(21,0.0000124406970836)
(22,0.0141586056698)
(23,0.000175322174347)
(24,0.0301540426166)
(25,0.158486300381)
(26,0.00411620690777)
(27,0.00000156029051995)
(28,0.00196247385779)
(29,0.0083424803803)
(30,0.00519733113672)
(31,0.00540525894775)
(32,0.000101911707235)
(33,0.0121564485318)
(34,0.00428134507692)
(35,0.00428134507692)
(36,0.00188485190586)
(37,0.00387724111164)
(38,0.0032708405131)
};
    \end{axis}
\end{tikzpicture}
\caption{Weights assigned to features by SVM}\label{fig:result_svm_weights}
\end{figure}
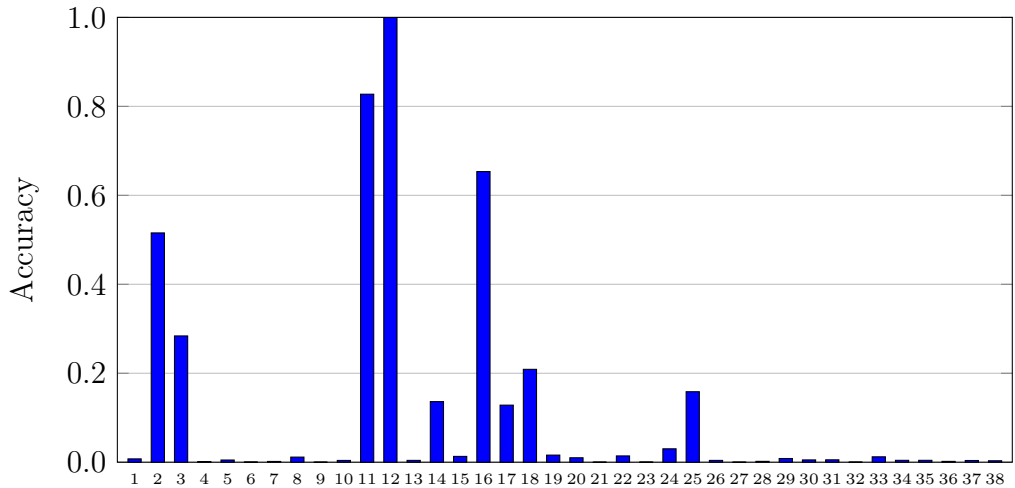

%% file: figures/SVM_RFE.tex
\begin{figure}[htbp]
\centering
\begin{tikzpicture}[scale=0.9]
\begin{axis}[width=0.75\textwidth,height=0.45\textwidth,
    		  xmin=1.0,xmax=38.0,
                   ymin=0.84,ymax=0.94,legend pos=south east,
                   y tick label style={
    			/pgf/number format/.cd,
   			fixed,
   			fixed zerofill,
    			precision=2},
                   xlabel={Number of selected features},ylabel={Cross validation score}] 
\addplot[color=blue,thick] coordinates {
(1,0.85693225)
(2,0.85673785)
(3,0.85701326)
(4,0.85728868)
(5,0.87775024)
(6,0.89349656)
(7,0.89767705)
(8,0.90480502)
(9,0.90397881)
(10,0.90402734)
(11,0.90315255)
(12,0.90164573)
(13,0.90115975)
(14,0.89963703)
(15,0.899799)
(16,0.89751473)
(17,0.89785493)
(18,0.89808173)
(19,0.89592719)
(20,0.89667237)
(21,0.89792009)
(22,0.89845448)
(23,0.89932931)
(24,0.90153252)
(25,0.90320111)
(26,0.90337942)
(27,0.90346039)
(28,0.90457819)
(29,0.90511276)
(30,0.90530715)
(31,0.90517755)
(32,0.90522614)
(33,0.90517754)
(34,0.90522613)
(35,0.90524233)
(36,0.91032968)
(37,0.91609743)
(38,0.92071416)
};
\end{axis}
\end{tikzpicture}
\vglue -0.05in
\caption{Recursive feature elimination with SVM}\label{fig:result_rfe}
\end{figure}
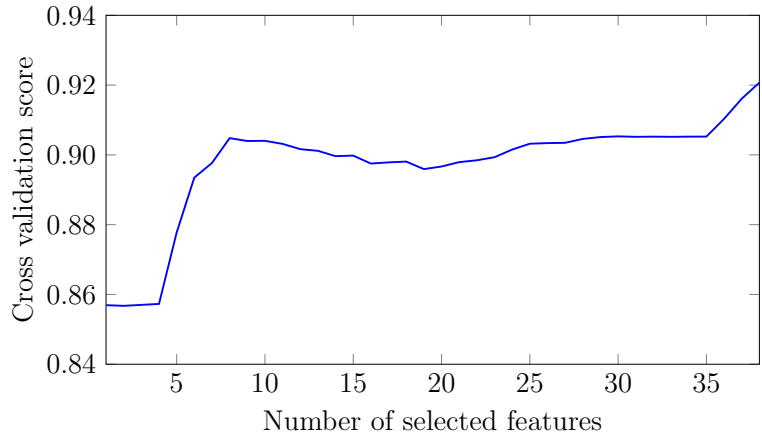

%% file: figures/RF_estimators.tex
\begin{figure}[!tbp]
\centering
\begin{tikzpicture}[scale=0.9]
\begin{axis}[width=0.75\textwidth,height=0.45\textwidth,
		  xmin=-4.0,xmax=1005.0,
                   ymin=0.96,ymax=1.00,legend pos=south east,
                   x tick label style={
    			/pgf/number format/.cd,
   			fixed,
   			fixed zerofill,
    			precision=0,
			1000 sep={}},
                   y tick label style={
    			/pgf/number format/.cd,
   			fixed,
   			fixed zerofill,
    			precision=2},
		  xtick = {1,100,200,300,400,500,600,700,800,900,1000},
		  ytick = {0.96,0.97,0.98,0.99,1.00},
                   xlabel={Number of estimators},ylabel={Cross validation score}] 
\addplot[color=blue,thick,mark=star,mark size=2.0] coordinates {
(1,0.970126102)
(2,0.968262989)
(3,0.982325111)
(4,0.982130746)
(5,0.984739014)
(6,0.985176468)
(7,0.98499823)
(8,0.985743488)
(9,0.986375319)
(10,0.986812733)
(20,0.987946788)
(30,0.988076374)
(40,0.988286979)
(50,0.988060134)
(100,0.988124909)
(200,0.988222087)
(300,0.988189666)
(400,0.988270712)
(500,0.988173529)
(600,0.988238307)
(700,0.988076301)
(800,0.988124873)
(900,0.988108674)
(1000,0.988141099)
};
\end{axis}
\end{tikzpicture}
\vglue -0.05in
\caption{Random forest accuracy as a function of the number of estimators}\label{fig:result_rf_estimators}
\end{figure}
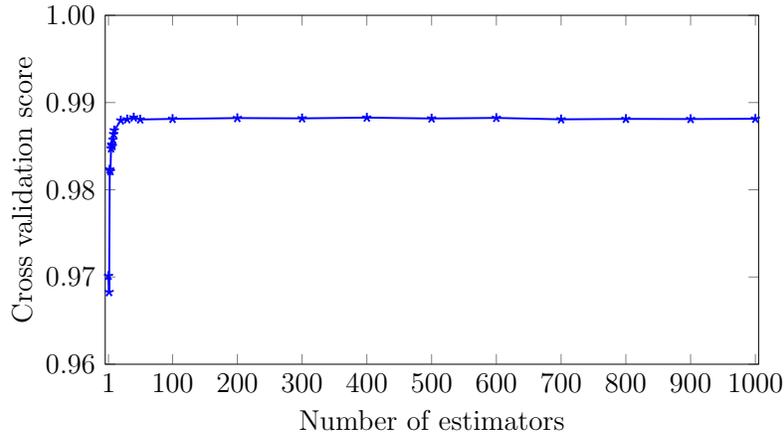

%% file: figures/RF_feature_importance.tex
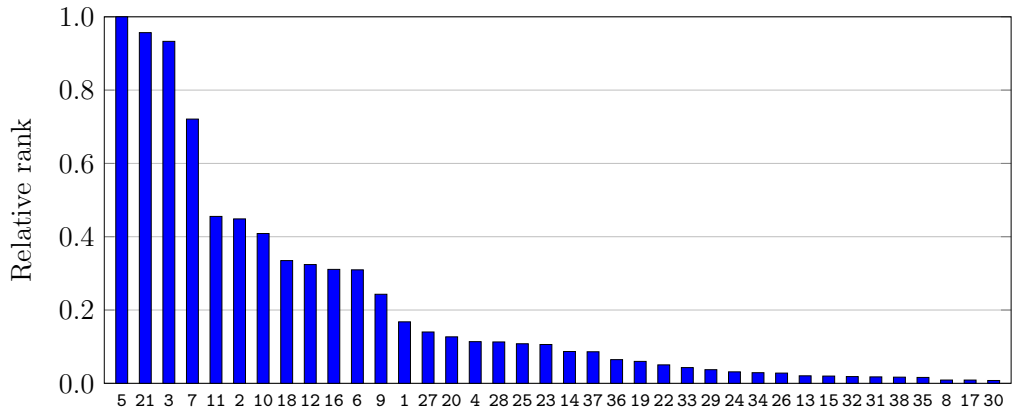
\begin{figure}[!tbp]
\centering
\begin{tikzpicture}[scale=0.9]
    \begin{axis}[
        width  = 1.0*\textwidth,
        height = 7.0cm,
        major x tick style = transparent,
        ybar=4*\pgflinewidth,
        bar width=5pt,
        ymajorgrids = true,
        ylabel = {Relative rank},
        symbolic x coords={%
5,
21,
3,
7,
11,
2,
10,
18,
12,
16,
6,
9,
1,
27,
20,
4,
28,
25,
23,
14,
37,
36,
19,
22,
33,
29,
24,
34,
26,
13,
15,
32,
31,
38,
35,
8,
17,
30
},
	y tick label style={
    	/pgf/number format/.cd,
   	fixed,
   	fixed zerofill,
    	precision=1},
        xtick = data,
        x tick label style={inner sep=0mm,font=\scriptsize\tt},
        scaled y ticks = false,
        enlarge x limits=0.02,
        ymin=0,
        ymax=1.0,
        legend cell align=left,
        legend pos=south east,
    ]
\addplot[fill=blue]
coordinates {
(5,1)
(21,0.956784416388)
(3,0.933021082506)
(7,0.720954318028)
(11,0.45539421862)
(2,0.448610756624)
(10,0.408718696483)
(18,0.334871555181)
(12,0.324033062381)
(16,0.310886554078)
(6,0.309647028835)
(9,0.243024120139)
(1,0.167781758594)
(27,0.140098567785)
(20,0.12683909547)
(4,0.11371146936)
(28,0.112931678268)
(25,0.107981128265)
(23,0.105957850261)
(14,0.0868234211433)
(37,0.0861339548092)
(36,0.0646552314344)
(19,0.0598641653391)
(22,0.0503709653364)
(33,0.0429549313698)
(29,0.0372238500256)
(24,0.031321118615)
(34,0.0290067938063)
(26,0.0279523482838)
(13,0.0205138831685)
(15,0.019811336186)
(32,0.0184211955603)
(31,0.0175946614751)
(38,0.0168483795842)
(35,0.0161032806006)
(8,0.00898747653926)
(17,0.00893064248691)
(30,0.00776929418136)
};
    \end{axis}
\end{tikzpicture}
\caption{Random forest RFE feature rank}\label{fig:result_rf_feature_imp}
\end{figure}

%% file: figures/RF_RFE.tex
\begin{figure}[!tbp]
\centering
\begin{tikzpicture}[scale=0.9]
\begin{axis}[width=0.75\textwidth,height=0.45\textwidth,
		  xmin=1.0,xmax=38.0,
                   ymin=0.92,ymax=0.99,legend pos=south east,
                   y tick label style={
    			/pgf/number format/.cd,
   			fixed,
   			fixed zerofill,
    			precision=2},
                   xlabel={Number of selected features},ylabel={Cross validation score}] 
\addplot[color=blue,thick] coordinates {
(1,0.92923564)
(2,0.9418235)
(3,0.95410361)
(4,0.97166511)
(5,0.9769789)
(6,0.98101288)
(7,0.98420433)
(8,0.98550041)
(9,0.98575965)
(10,0.98593784)
(11,0.98601884)
(12,0.98656965)
(13,0.98721766)
(14,0.98746067)
(15,0.9875255)
(16,0.98762269)
(17,0.98775229)
(18,0.98789809)
(19,0.98776849)
(20,0.98771988)
(21,0.98771988)
(22,0.98760648)
(23,0.98762269)
(24,0.98775228)
(25,0.98784949)
(26,0.9878981)
(27,0.98791429)
(28,0.98797909)
(29,0.98783329)
(30,0.9878981)
(31,0.98799529)
(32,0.98788189)
(33,0.98784949)
(34,0.98760648)
(35,0.98801149)
(36,0.98783331)
(37,0.9881087)
(38,0.9878981)
};
\end{axis}
\end{tikzpicture}
\vglue -0.05in
\caption{RFE results for random forest}\label{fig:result_rf_rfe}
\end{figure}
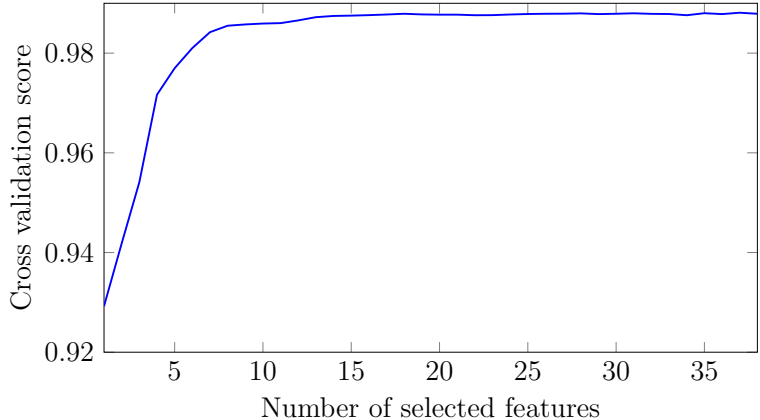

%% file: figures/XG_estimators.tex
\begin{figure}[!tbp]
\centering
\begin{tikzpicture}[scale=0.9]
\begin{axis}[width=0.75\textwidth,height=0.45\textwidth,
		  xmin=-4.0,xmax=2005.0,
                   ymin=0.93,ymax=1.00,legend pos=south east,
                   x tick label style={
    			/pgf/number format/.cd,
   			fixed,
   			fixed zerofill,
    			precision=0,
			1000 sep={}},
                   y tick label style={
    			/pgf/number format/.cd,
   			fixed,
   			fixed zerofill,
    			precision=2},
		  xtick = {1,200,400,600,800,1000,1200,1400,1600,1800,2000},
		  ytick = {0.93,0.94,0.95,0.96,0.97,0.98,0.99,1.00},
                   xlabel={Number of estimators},ylabel={Cross validation score}] 
\addplot[color=blue,thick,mark=star,mark size=2.0] coordinates {
(1,0.939182737)
(2,0.939831071)
(3,0.942715049)
(4,0.944869467)
(5,0.951689656)
(6,0.954589439)
(7,0.954508427)
(8,0.95510787)
(9,0.955383291)
(10,0.955853076)
(20,0.95995191)
(30,0.963791548)
(40,0.966707663)
(50,0.972215932)
(100,0.972296961)
(200,0.979295635)
(300,0.985046897)
(400,0.987946813)
(500,0.989226664)
(600,0.990117695)
(700,0.9906847)
(800,0.991041125)
(900,0.991267936)
(1000,0.991365127)
(1100,0.991656738)
(1200,0.991672948)
(1300,0.991705343)
(1400,0.991867346)
(1500,0.991753947)
(1600,0.991834953)
(1700,0.991915948)
(1800,0.991980756)
(1900,0.992029353)
(2000,0.992013146)
};
\end{axis}
\end{tikzpicture}
\vglue -0.05in
\caption{XGBoost accuracy as a function of the number of estimators}\label{fig:result_xgboost_estimators}
\end{figure}
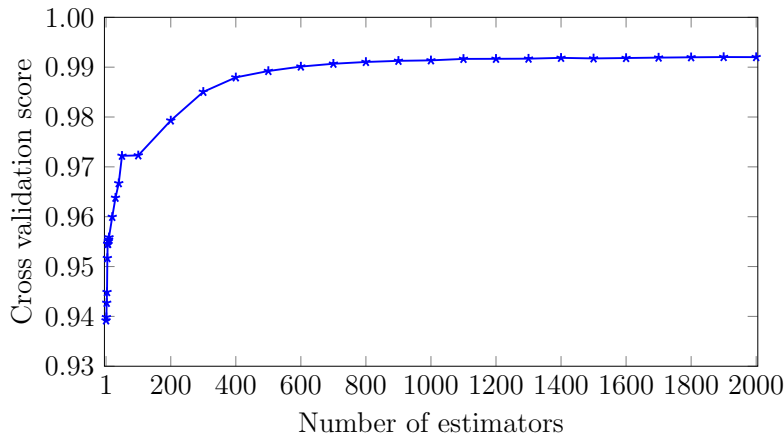

%% file: figures/XG_feature_importance.tex
\begin{figure}[!tbp]
\centering
\begin{tikzpicture}[scale=0.9]
    \begin{axis}[
        width  = 1.0*\textwidth,
        height = 7.5cm,
        major x tick style = transparent,
        ybar=4*\pgflinewidth,
        bar width=5pt,
        ymajorgrids = true,
        ylabel = {Relative rank},
        symbolic x coords={%
2,
21,
5,
6,
7,
10,
9,
3,
16,
20,
11,
22,
23,
1,
12,
18,
36,
28,
4,
25,
27,
8,
29,
14,
37,
24,
15,
13,
32,
26,
30,
34,
17,
19,
31,
33,
35,
38
},
	y tick label style={
    	/pgf/number format/.cd,
   	fixed,
   	fixed zerofill,
    	precision=1},
        xtick = data,
        x tick label style={inner sep=0mm,font=\scriptsize\tt},
        scaled y ticks = false,
        enlarge x limits=0.02,
        ymin=0,
        ymax=1.0,
        legend cell align=left,
        legend pos=south east,
    ]
\addplot[fill=blue]
coordinates {
(2,1)
(21,0.910447716713)
(5,0.874626874924)
(6,0.774626851082)
(7,0.686567187309)
(10,0.650746226311)
(9,0.640298485756)
(3,0.523880600929)
(16,0.462686568499)
(20,0.304477602243)
(11,0.285074621439)
(22,0.194029837847)
(23,0.15373134613)
(1,0.117910444736)
(12,0.110447756946)
(18,0.0985074564815)
(36,0.0805970132351)
(28,0.0656716376543)
(4,0.0597014911473)
(25,0.0582089535892)
(27,0.053731340915)
(8,0.0492537282407)
(29,0.0328358188272)
(14,0.0298507455736)
(37,0.0268656704575)
(24,0.0179104469717)
(15,0.0149253727868)
(13,0.0134328352287)
(32,0.0119402976707)
(26,0.0104477610439)
(30,0.00597014883533)
(34,0.00298507441767)
(17,0.00149253720883)
(19,0.00149253720883)
(31,0.00149253720883)
(33,0.00149253720883)
(35,0)
(38,0)
};
    \end{axis}
\end{tikzpicture}
\caption{XGBoost feature importance}\label{fig:result_xgboost_feature_imp}
\end{figure}
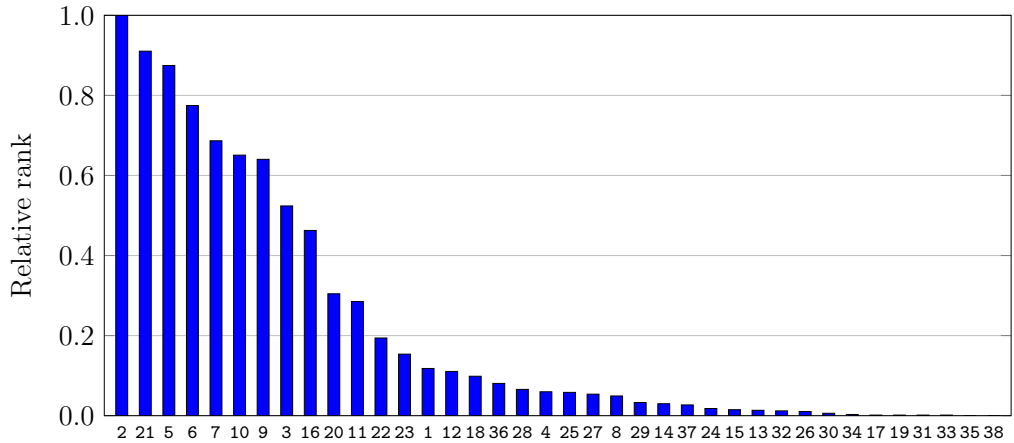

%% file: figures/XG_RFE.tex
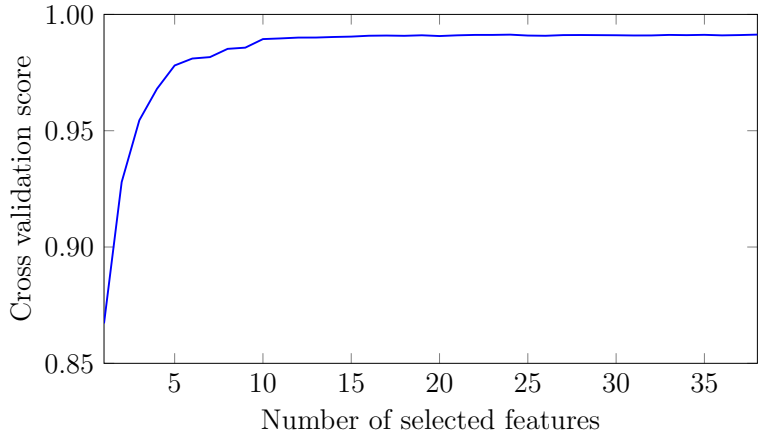
\begin{figure}[!tbp]
\centering
\begin{tikzpicture}[scale=0.9]
\begin{axis}[width=0.75\textwidth,height=0.45\textwidth,
		  xmin=1.0,xmax=38.0,
                   ymin=0.85,ymax=1.00,legend pos=south east,
                   y tick label style={
    			/pgf/number format/.cd,
   			fixed,
   			fixed zerofill,
    			precision=2},
                   xlabel={Number of selected features},ylabel={Cross validation score}] 
\addplot[color=blue,thick] coordinates {
(1,0.867171)
(2,0.9279234)
(3,0.95449245)
(4,0.96801993)
(5,0.97804833)
(6,0.98102912)
(7,0.98164471)
(8,0.98519263)
(9,0.98571104)
(10,0.98938859)
(11,0.9896802)
(12,0.99003662)
(13,0.99005281)
(14,0.99029581)
(15,0.99045782)
(16,0.99083045)
(17,0.99091145)
(18,0.99079806)
(19,0.99102485)
(20,0.99070084)
(21,0.99100865)
(22,0.99118686)
(23,0.99118686)
(24,0.99130025)
(25,0.99091143)
(26,0.99081423)
(27,0.99110584)
(28,0.99113825)
(29,0.99108964)
(30,0.99104106)
(31,0.99094385)
(32,0.99096004)
(33,0.99118686)
(34,0.99108965)
(35,0.99121926)
(36,0.99099245)
(37,0.99112206)
(38,0.99131646)
};
\end{axis}
\end{tikzpicture}
\vglue -0.05in
\caption{Recursive feature elimination with XGBoost}\label{fig:result_xgboost_rfe}
\end{figure}

%% file: figures/ML_comparison.tex
\begin{figure}[!tb]
\centering
\begin{tikzpicture}[scale=0.9]
    \begin{axis}[
        width  = 0.625*\textwidth,
        height = 0.5*\textwidth,
        major x tick style = transparent,
        ybar=4*\pgflinewidth,
        bar width=25pt,
        ymajorgrids = true,
        ylabel = {Accuracy},
        symbolic x coords={SVM,Random Forest,XGBoost},
	y tick label style={
    	/pgf/number format/.cd,
   	fixed,
   	fixed zerofill,
    	precision=2},
        xtick = data,
        x tick label style={rotate=45,anchor=north east, inner sep=0mm},
        scaled y ticks = false,
        enlarge x limits=0.25,
        ymin=0.8,
        ymax=1.0,
        legend cell align=left,
        legend pos=south east,
    ]
        \addplot[fill=blue]
            coordinates {
(SVM,0.920454966)
(Random Forest,0.988141158)
(XGBoost,0.991510834)
};
    \end{axis}
\end{tikzpicture}
\vglue-0.05in
\caption{Accuracy comparison of machine learning algorithms}\label{fig:result_svm_rf_xgb}
\end{figure}
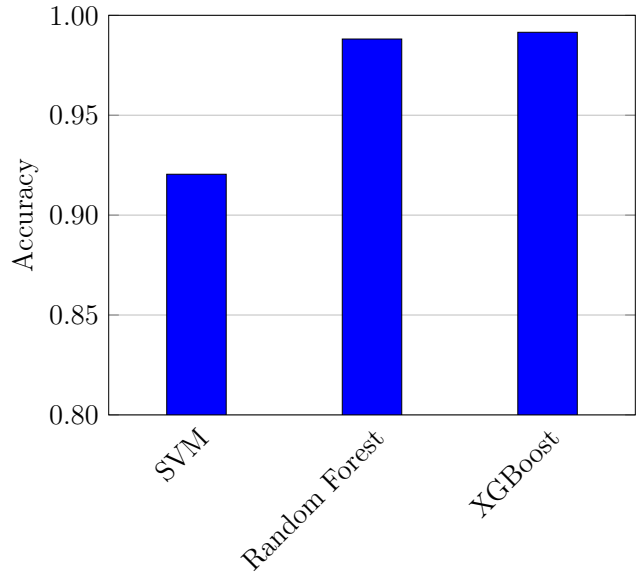

%% file: figures/classification.tex
\begin{figure}[!tb]
\centering
\begin{tikzpicture}[scale=0.9]
    \begin{axis}[
        width  = 0.75*\textwidth,
        height = 0.5*\textwidth,
        major x tick style = transparent,
        ybar=4*\pgflinewidth,
        bar width=8pt,
        ymajorgrids = true,
        ylabel = {Accuracy},
        symbolic x coords={Dridex vs Trickbot,Dridex vs WannaCry,Dridex vs Zbot,Trickbot vs WannaCry,Trickbot vs Zbot,WannaCry vs Zbot},
	y tick label style={
    	/pgf/number format/.cd,
   	fixed,
   	fixed zerofill,
    	precision=2},
        xtick = data,
        x tick label style={rotate=45,anchor=north east, inner sep=0mm},
        scaled y ticks = false,
        enlarge x limits=0.12,
        ymin=0.75,
        ymax=1.0,
        legend cell align=left,
        legend pos=south east,
    ]
\addplot[fill=red]
coordinates {
(Dridex vs Trickbot,0.946736597)
(Dridex vs WannaCry,1)
(Dridex vs Zbot,1)
(Trickbot vs WannaCry,1)
(Trickbot vs Zbot,0.984450142)
(WannaCry vs Zbot,1)
};
\addplot[fill=green]
coordinates {
(Dridex vs Trickbot,0.995652174)
(Dridex vs WannaCry,1)
(Dridex vs Zbot,1)
(Trickbot vs WannaCry,1)
(Trickbot vs Zbot,0.992296296)
(WannaCry vs Zbot,1)
};
\addplot[fill=blue]
coordinates {
(Dridex vs Trickbot,0.979166667)
(Dridex vs WannaCry,0.963333333)
(Dridex vs Zbot,0.985714286)
(Trickbot vs WannaCry,0.992153846)
(Trickbot vs Zbot,0.980769231)
(WannaCry vs Zbot,1)
};
\legend{SVM,Random Forest,XGBoost}
\end{axis}
\end{tikzpicture}
\vglue-0.05in
\caption{Accuracy for pairwise classification}\label{fig:result_multiclass}
\end{figure}
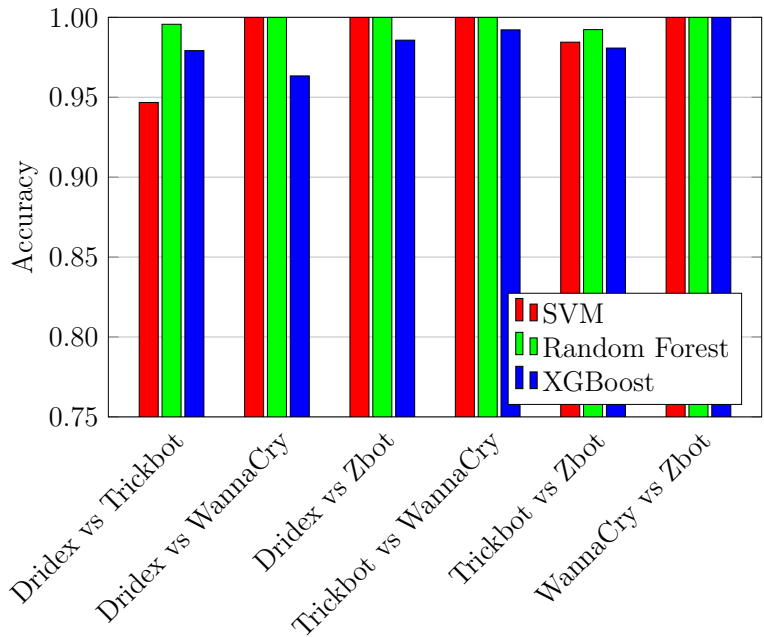

%% file: figures/multiclass.tex
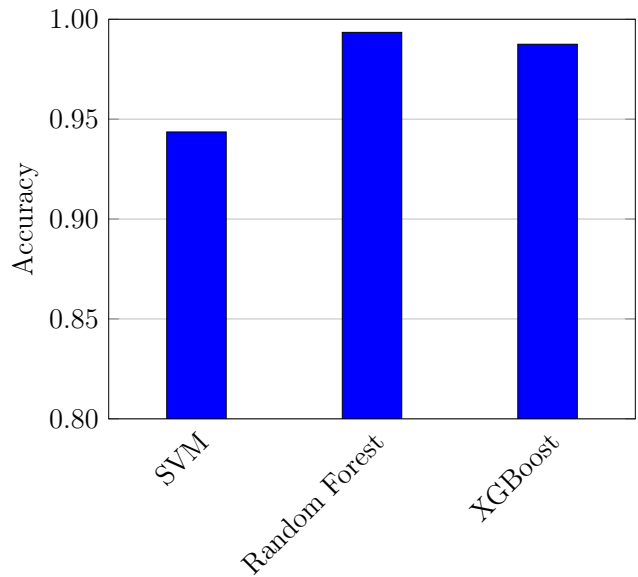
\begin{figure}[!tbp]
\centering
\begin{tikzpicture}[scale=0.9]
    \begin{axis}[
        width  = 0.625*\textwidth,
        height = 0.5*\textwidth,
        major x tick style = transparent,
        ybar=4*\pgflinewidth,
        bar width=25pt,
        ymajorgrids = true,
        ylabel = {Accuracy},
        symbolic x coords={SVM,Random Forest,XGBoost},
	y tick label style={
    	/pgf/number format/.cd,
   	fixed,
   	fixed zerofill,
    	precision=2},
        xtick = data,
        x tick label style={rotate=45,anchor=north east, inner sep=0mm},
        scaled y ticks = false,
        enlarge x limits=0.25,
        ymin=0.8,
        ymax=1.0,
        legend cell align=left,
        legend pos=south east,
    ]
        \addplot[fill=blue]
            coordinates {
(SVM,0.943591872)
(Random Forest,0.99344086)
(XGBoost,0.9875)
};
    \end{axis}
\end{tikzpicture}
\vglue-0.05in
\caption{Accuracy for the multiclass problem}\label{fig:result_multiclass_all}
\end{figure}

%% file: figures/measures_compared.tex
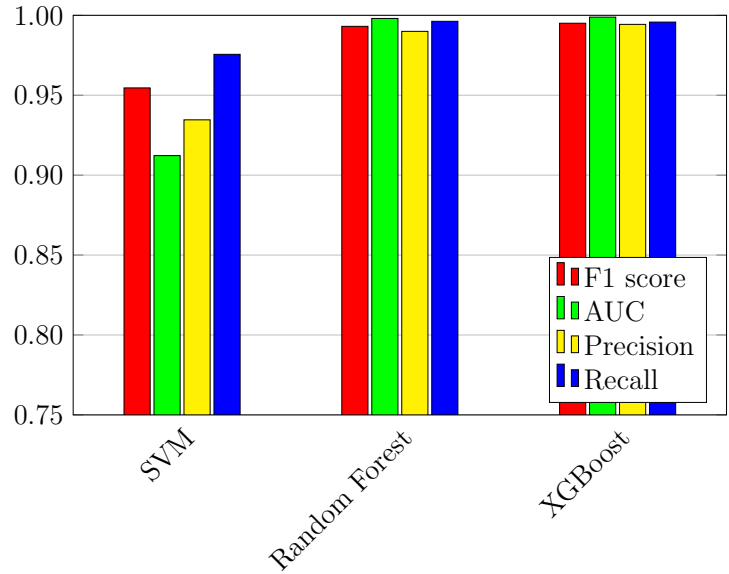
\begin{figure}[!tbp]
\centering
\begin{tikzpicture}[scale=0.9]
    \begin{axis}[
        width  = 0.75*\textwidth,
        height = 0.5*\textwidth,
        major x tick style = transparent,
        ybar=4*\pgflinewidth,
        bar width=11pt,
        ymajorgrids = true,
        symbolic x coords={SVM,Random Forest,XGBoost},
	y tick label style={
    	/pgf/number format/.cd,
   	fixed,
   	fixed zerofill,
    	precision=2},
        xtick = data,
        x tick label style={rotate=45,anchor=north east, inner sep=0mm},
        scaled y ticks = false,
        enlarge x limits=0.25,
        ymin=0.75,
        ymax=1.0,
        legend cell align=left,
        legend pos=south east,
    ]
\addplot[fill=red]
coordinates {
(SVM,0.954582661)
(Random Forest,0.993103175)
(XGBoost,0.995051681)
};
\addplot[fill=green]
coordinates {
(SVM,0.912227037)
(Random Forest,0.998039123)
(XGBoost,0.998866439)
};
\addplot[fill=yellow]
coordinates {
(SVM,0.934605752)
(Random Forest,0.989970562)
(XGBoost,0.99433712)
};
\addplot[fill=blue]
coordinates {
(SVM,0.97559425)
(Random Forest,0.996256956)
(XGBoost,0.995768002)
};
\legend{F1 score,AUC,Precision,Recall}
\end{axis}
\end{tikzpicture}
\vglue-0.05in
\caption{Comparison based on various measures of success}\label{fig:ml_comparison}
\end{figure}

%% file: figures/heatmap.tex
\pgfkeys{
    /pgf/number format/int trunc, 
    /pgf/number format/fixed zerofill=true }
    
\pgfplotstableset{
    /color cells/min/.initial=0,
    /color cells/max/.initial=1000,
    /color cells/textcolor/.initial=,
    %
    color cells/.code={%
        \pgfqkeys{/color cells}{#1}%
        \pgfkeysalso{%
            postproc cell content/.code={%
                \begingroup
                %
                \pgfkeysgetvalue{/pgfplots/table/@preprocessed cell content}\value
\ifx\value\empty
\endgroup
\else
                \pgfmathfloatparsenumber{\value}%
                \pgfmathfloattofixed{\pgfmathresult}%
                \let\value=\pgfmathresult
                %
                \pgfplotscolormapaccess
                    [\pgfkeysvalueof{/color cells/min}:\pgfkeysvalueof{/color cells/max}]%
                    {\value}%
                    {\pgfkeysvalueof{/pgfplots/colormap name}}%
                %
                \pgfkeysgetvalue{/pgfplots/table/@cell content}\typesetvalue
                \pgfkeysgetvalue{/color cells/textcolor}\textcolorvalue
                %
                \toks0=\expandafter{\typesetvalue}%
                \xdef\temp{%
                    \noexpand\pgfkeysalso{%
                        @cell content={%
                            \noexpand\cellcolor[rgb]{\pgfmathresult}%
                            \noexpand\definecolor{mapped color}{rgb}{\pgfmathresult}%
                            \ifx\textcolorvalue\empty
                            \else
                                \noexpand\color{\textcolorvalue}%
                            \fi
                        }%
                    }%
                }%
                \endgroup
                \temp
\fi
            }%
        }%
    }
}
\begin{figure}[!tbp]
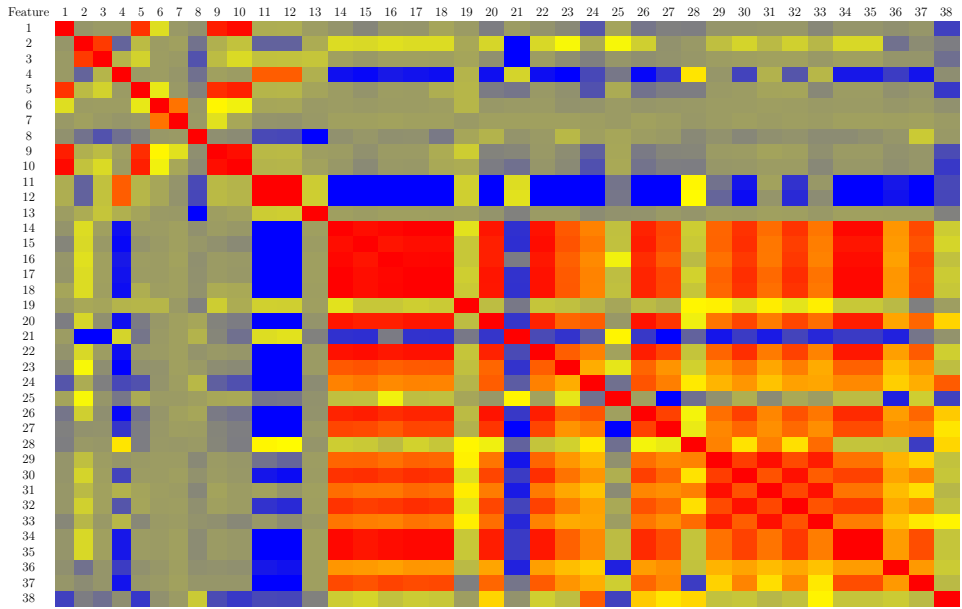

\begin{adjustbox}{width=0.85\textwidth,center}
\pgfplotstabletypeset[%
     color cells={min=-0.25,max=1.0,textcolor=black},
    col sep=comma,
    columns/Feature/.style={reset styles,string type}%
]{
Feature,1,2,3,4,5,6,7,8,9,10,11,12,13,14,15,16,17,18,19,20,21,22,23,24,25,26,27,28,29,30,31,32,33,34,35,36,37,38 
1,1,-0.00516,0.000962,0.00093,0.831358,0.111767,-0.000871,-0.013436,0.904227,0.967217,0.032122,0.033223,0.005844,-0.00791,-0.033505,-0.006079,-0.007649,0.018805,0.003312,-0.045434,0.004801,-0.010022,-0.027332,-0.111561,0.018169,-0.058828,-0.041189,-0.044193,-0.007643,-0.007526,-0.003192,-0.003531,-0.030075,-0.006794,-0.006794,-0.01536,-0.005263,-0.145099
2,-0.00516,1,0.807432,-0.087547,0.055337,0.004796,0.011653,-0.063596,0.037648,0.065646,-0.089697,-0.090644,0.030799,0.108031,0.103953,0.115703,0.108164,0.108219,0.02399,0.09955,-0.317339,0.103011,0.154753,0.031006,0.153091,0.08759,-0.01228,-0.000946,0.061138,0.095018,0.053867,0.089042,0.048454,0.10231,0.10231,-0.065609,-0.021884,-0.045935
3,0.000962,0.807432,1,0.045427,0.092211,0.006465,0.001026,-0.11565,0.056829,0.105126,0.065473,0.066966,0.070535,0.010463,0.000257,0.010534,0.010288,0.009935,0.019029,-0.015055,-0.487301,0.005427,-0.017894,-0.043569,-0.004897,-0.017508,-0.010282,-0.004128,0.006284,0.008561,0.013356,0.012682,-0.000991,0.010734,0.010734,-0.006744,-0.006549,-0.069497
4,0.00093,-0.087547,0.045427,1,0.002611,0.00602,-0.005237,-0.06812,0.010922,0.004768,0.69427,0.69258,0.037775,-0.225676,-0.237919,-0.210268,-0.221307,-0.228533,0.043379,-0.225815,0.095814,-0.228044,-0.421859,-0.132119,-0.055159,-0.240482,-0.164274,0.250496,-0.005174,-0.140609,0.039493,-0.111292,0.048208,-0.212162,-0.212162,-0.151705,-0.218591,-0.01768
5,0.831358,0.055337,0.092211,0.002611,1,0.128084,-0.001146,-0.037559,0.850361,0.896059,0.045269,0.046817,0.018243,0.004972,-0.00912,0.005435,0.004738,0.032951,0.04615,-0.049744,-0.05852,-0.0001,-0.014229,-0.118276,0.028988,-0.063485,-0.046377,-0.046451,0.001092,0.004261,0.016638,0.019015,-0.028307,0.004658,0.004658,-0.01355,-0.004029,-0.161048
6,0.111767,0.004796,0.006465,0.00602,0.128084,1,0.620912,-0.001648,0.196738,0.142227,0.021415,0.022042,0.000597,0.004872,0.000799,0.004073,0.003768,0.007941,0.046714,-0.002694,0.001301,0.002487,-0.010556,-0.008841,0.001988,-0.004737,-0.002079,-0.001084,0.003862,0.00473,0.007661,0.006141,0.001747,0.003543,0.003543,-0.003424,0.000505,-0.014082
7,-0.000871,0.011653,0.001026,-0.005237,-0.001146,0.620912,1,0.001925,0.116245,0.021931,-0.009367,-0.009562,-0.001054,0.01103,0.010824,0.013666,0.010933,0.011003,0.000347,0.011217,0.011289,0.010739,0.011189,0.008402,0.002326,0.012246,0.008337,-0.001987,0.00479,0.009384,0.003636,0.008205,0.004275,0.010164,0.010164,0.009998,0.012559,0.003141
8,-0.013436,-0.063596,-0.11565,-0.06812,-0.037559,-0.001648,0.001925,1,-0.017509,-0.023239,-0.130792,-0.133451,-0.56603,-0.012879,-0.004479,-0.014744,-0.013518,-0.05188,0.02144,0.04177,-0.009194,0.000994,0.050023,0.00959,0.020268,0.006359,0.000619,-0.027093,-0.011743,-0.030493,-0.015814,-0.009715,-0.021461,-0.021461,-0.015284,-0.004946,0.079689,0.0
9,0.904227,0.037648,0.056829,0.010922,0.850361,0.196738,0.116245,-0.017509,1,0.955906,0.051946,0.053523,0.007501,0.004016,-0.014189,0.003094,0.002452,0.029751,0.084851,-0.036132,-0.031706,-0.00085,-0.020439,-0.095124,0.021643,-0.051201,-0.033657,-0.030973,0.003115,0.004174,0.015291,0.010974,-0.015428,0.002915,0.002915,-0.012075,-0.005468,-0.127691
10,0.967217,0.065646,0.105126,0.004768,0.896059,0.142227,0.021931,-0.023239,0.955906,1,0.044325,0.045771,0.014292,-0.002378,-0.024394,-0.001182,-0.002597,0.026207,0.030874,-0.04758,-0.06771,-0.006002,-0.024322,-0.118915,0.025756,-0.062397,-0.044353,-0.045553,-0.003595,-0.00257,0.005191,0.003849,-0.027526,-0.002377,-0.002377,-0.015143,-0.006145,-0.15853
11,0.032122,-0.089697,0.065473,0.69427,0.045269,0.021415,-0.009367,-0.130792,0.051946,0.044325,1,0.996786,0.083025,-0.310854,-0.335335,-0.294153,-0.306144,-0.314325,0.089872,-0.338359,0.111077,-0.318272,-0.539122,-0.276877,-0.060422,-0.3611,-0.250845,0.196611,-0.064272,-0.216382,0.014454,-0.168644,-0.001545,-0.293843,-0.293843,-0.211846,-0.305124,-0.131942
12,0.033223,-0.090644,0.066966,0.69258,0.046817,0.022042,-0.009562,-0.133451,0.053523,0.045771,0.996786,1,0.085157,-0.321597,-0.346724,-0.303517,-0.317279,-0.324971,0.088215,-0.349585,0.118442,-0.328721,-0.553896,-0.286221,-0.058339,-0.372861,-0.261621,0.18032,-0.085409,-0.231641,-0.006922,-0.183861,-0.022371,-0.303698,-0.303698,-0.222402,-0.308339,-0.136032
13,0.005844,0.030799,0.070535,0.037775,0.018243,0.000597,-0.001054,-0.56603,0.007501,0.014292,0.083025,0.085157,1,0.008673,0.004658,0.00916,0.008766,0.00837,0.009488,-0.008394,-0.038885,0.003168,0.003509,-0.022404,-0.011662,-0.010773,0.000926,-0.007469,0.005662,0.003833,0.01022,0.009599,-0.000038,0.0108,0.0108,0.008821,0.00991,-0.039037
14,-0.00791,0.108031,0.010463,-0.225676,0.004972,0.004872,0.01103,-0.012879,0.004016,-0.002378,-0.310854,-0.321597,0.008673,1,0.955106,0.979519,0.997492,0.998655,0.117808,0.928121,-0.171772,0.941953,0.706395,0.572003,0.062338,0.882049,0.77027,0.083888,0.668958,0.834225,0.64361,0.832065,0.61414,0.975167,0.975167,0.507749,0.759049,0.082222
15,-0.033505,0.103953,0.000257,-0.237919,-0.00912,0.000799,0.010824,-0.004479,-0.014189,-0.024394,-0.335335,-0.346724,0.004658,0.955106,1,0.933672,0.954033,0.954997,0.074003,0.895354,-0.175063,0.957144,0.727611,0.604741,0.064522,0.903845,0.756853,0.077339,0.672719,0.846677,0.609937,0.794777,0.585974,0.929768,0.929768,0.496464,0.726711,0.097781
16,-0.006079,0.115703,0.010534,-0.210268,0.005435,0.004073,0.013666,-0.014744,0.003094,-0.001182,-0.294153,-0.303517,0.00916,0.979519,0.933672,1,0.979203,0.979981,0.07385,0.895323,-0.048796,0.923277,0.685237,0.547832,0.143474,0.851546,0.700499,0.054604,0.629181,0.81306,0.605018,0.813845,0.57426,0.958114,0.958114,0.472384,0.777732,0.06402
17,-0.007649,0.108164,0.010288,-0.221307,0.004738,0.003768,0.010933,-0.013518,0.002452,-0.002597,-0.306144,-0.317279,0.008766,0.997492,0.954033,0.979203,1,0.997509,0.086117,0.927617,-0.169141,0.941429,0.705239,0.570379,0.059679,0.880551,0.771386,0.091901,0.678145,0.848389,0.655225,0.850193,0.623466,0.974029,0.974029,0.507257,0.753193,0.080757
18,0.018805,0.108219,0.009935,-0.228533,0.032951,0.007941,0.011003,-0.012038,0.029751,0.026207,-0.314325,-0.324971,0.00837,0.998655,0.954997,0.979981,0.997509,1,0.079915,0.926643,-0.169561,0.942409,0.706117,0.566717,0.063404,0.879667,0.768988,0.073606,0.662285,0.832601,0.636719,0.830526,0.6053,0.975561,0.975561,0.505589,0.761349,0.074756
19,0.003312,0.02399,0.019029,0.043379,0.04615,0.046714,0.000347,-0.039842,0.084851,0.030874,0.089872,0.088215,0.009488,0.117808,0.074003,0.07385,0.086117,0.079915,1,0.057991,-0.054405,0.070524,0.063176,0.044627,0.020828,0.054908,0.039738,0.195564,0.21103,0.114037,0.224393,0.121219,0.218002,0.076645,0.076645,0.052695,-0.042211,0.007703
20,-0.045434,0.09955,-0.015055,-0.225815,-0.049744,-0.002694,0.011217,0.02144,-0.036132,-0.04758,-0.338359,-0.349585,-0.008394,0.928121,0.895354,0.895323,0.927617,0.926643,0.057991,1,-0.163132,0.885061,0.663265,0.693422,-0.002002,0.934137,0.823673,0.14138,0.616773,0.772429,0.592803,0.769763,0.638798,0.904219,0.904219,0.460376,0.664939,0.304604
21,0.004801,-0.317339,-0.487301,0.095814,-0.05852,0.001301,0.011289,0.04177,-0.031706,-0.06771,0.111077,0.118442,-0.038885,-0.171772,-0.175063,-0.048796,-0.169141,-0.169561,-0.054405,-0.163132,1,-0.129755,-0.166885,-0.09941,0.198254,-0.158815,-0.247701,-0.090033,-0.215706,-0.15311,-0.208349,-0.138397,-0.186221,-0.155044,-0.155044,-0.196059,-0.059611,-0.009963
22,-0.010022,0.103011,0.005427,-0.228044,-0.0001,0.002487,0.010739,-0.009194,-0.00085,-0.006002,-0.318272,-0.328721,0.003168,0.941953,0.957144,0.923277,0.941429,0.942409,0.070524,0.885061,-0.129755,1,0.692211,0.577181,0.014296,0.904264,0.772088,0.06982,0.667474,0.846476,0.59797,0.78498,0.57138,0.921675,0.921675,0.479462,0.699368,0.087176
23,-0.027332,0.154753,-0.017894,-0.421859,-0.014229,-0.010556,0.011189,0.000994,-0.020439,-0.024322,-0.539122,-0.553896,0.003509,0.706395,0.727611,0.685237,0.705239,0.706117,0.063176,0.663265,-0.166885,0.692211,1,0.433072,0.125725,0.669711,0.515732,0.091618,0.504518,0.622407,0.451978,0.587035,0.438655,0.681005,0.681005,0.379317,0.511363,0.060575
24,-0.111561,0.031006,-0.043569,-0.132119,-0.118276,-0.008841,0.008402,0.050023,-0.095124,-0.118915,-0.276877,-0.286221,-0.022404,0.572003,0.604741,0.547832,0.570379,0.566717,0.044627,0.693422,-0.09941,0.577181,0.433072,1,-0.067024,0.723202,0.58353,0.235398,0.408702,0.507651,0.361614,0.473754,0.454592,0.549442,0.549442,0.322967,0.42937,0.699171
25,0.018169,0.153091,-0.004897,-0.055159,0.028988,0.001988,0.002326,0.00959,0.021643,0.025756,-0.060422,-0.058339,-0.011662,0.062338,0.064522,0.143474,0.059679,0.063404,0.020828,-0.002002,0.198254,0.014296,0.125725,-0.067024,1,0.010305,-0.531091,-0.069607,-0.014032,0.034961,-0.023052,0.026161,0.005529,0.056679,0.056679,-0.197464,0.08545,-0.143793
26,-0.058828,0.08759,-0.017508,-0.240482,-0.063485,-0.004737,0.012246,0.020268,-0.051201,-0.062397,-0.3611,-0.372861,-0.010773,0.882049,0.903845,0.851546,0.880551,0.879667,0.054908,0.934137,-0.158815,0.904264,0.669711,0.723202,0.010305,1,0.785462,0.148452,0.634267,0.792511,0.560373,0.729505,0.61011,0.859245,0.859245,0.484788,0.664343,0.338659
27,-0.041189,-0.01228,-0.010282,-0.164274,-0.046377,-0.002079,0.008337,0.006359,-0.033657,-0.044353,-0.250845,-0.261621,0.000926,0.77027,0.756853,0.700499,0.771386,0.768988,0.039738,0.823673,-0.247701,0.772088,0.515732,0.58353,-0.531091,0.785462,1,0.131738,0.555742,0.665995,0.531638,0.65548,0.534696,0.752978,0.752978,0.535293,0.560471,0.250467
28,-0.044193,-0.000946,-0.004128,0.250496,-0.046451,-0.001084,-0.001987,0.000619,-0.030973,-0.045553,0.196611,0.18032,-0.007469,0.083888,0.077339,0.054604,0.091901,0.073606,0.195564,0.14138,-0.090033,0.06982,0.091618,0.235398,-0.069607,0.148452,0.131738,1,0.575036,0.25615,0.581344,0.267184,0.649205,0.069373,0.069373,0.066835,-0.152255,0.299463
29,-0.007643,0.061138,0.006284,-0.005174,0.001092,0.003862,0.00479,-0.027093,0.003115,-0.003595,-0.064272,-0.085409,0.005662,0.668958,0.672719,0.629181,0.678145,0.662285,0.21103,0.616773,-0.215706,0.667474,0.504518,0.408702,-0.014032,0.634267,0.555742,0.575036,1,0.799514,0.952474,0.755606,0.905401,0.624719,0.624719,0.409872,0.311463,0.065803
30,-0.007526,0.095018,0.008561,-0.140609,0.004261,0.00473,0.009384,-0.011743,0.004174,-0.00257,-0.216382,-0.231641,0.003833,0.834225,0.846677,0.81306,0.848389,0.832601,0.114037,0.772429,-0.15311,0.846476,0.622407,0.507651,0.034961,0.792511,0.665995,0.25615,0.799514,1,0.730112,0.946213,0.693538,0.786817,0.786817,0.461163,0.565838,0.072468
31,-0.003192,0.053867,0.013356,0.039493,0.016638,0.007661,0.003636,-0.030493,0.015291,0.005191,0.014454,-0.006922,0.01022,0.64361,0.609937,0.605018,0.655225,0.636719,0.224393,0.592803,-0.208349,0.59797,0.451978,0.361614,-0.023052,0.560373,0.531638,0.581344,0.952474,0.730112,1,0.767025,0.945797,0.619295,0.619295,0.385183,0.271888,0.047944
32,-0.003531,0.089042,0.012682,-0.111292,0.019015,0.006141,0.008205,-0.015814,0.010974,0.003849,-0.168644,-0.183861,0.009599,0.832065,0.794777,0.813845,0.850193,0.830526,0.121219,0.769763,-0.138397,0.78498,0.587035,0.473754,0.026161,0.729505,0.65548,0.267184,0.755606,0.946213,0.767025,1,0.725922,0.810363,0.810363,0.448213,0.554328,0.064992
33,-0.030075,0.048454,-0.000991,0.048208,-0.028307,0.001747,0.004275,-0.009715,-0.015428,-0.027526,-0.001545,-0.022371,-0.000038,0.61414,0.585974,0.57426,0.623466,0.6053,0.218002,0.638798,-0.186221,0.57138,0.438655,0.454592,0.005529,0.61011,0.534696,0.649205,0.905401,0.693538,0.945797,0.725922,1,0.588373,0.588373,0.363722,0.238706,0.200552
34,-0.006794,0.10231,0.010734,-0.212162,0.004658,0.003543,0.010164,-0.021461,0.002915,-0.002377,-0.293843,-0.303698,0.0108,0.975167,0.929768,0.958114,0.974029,0.975561,0.076645,0.904219,-0.155044,0.921675,0.681005,0.549442,0.056679,0.859245,0.752978,0.069373,0.624719,0.786817,0.619295,0.810363,0.588373,1,1,0.487852,0.746433,0.071418
35,-0.006794,0.10231,0.010734,-0.212162,0.004658,0.003543,0.010164,-0.021461,0.002915,-0.002377,-0.293843,-0.303698,0.0108,0.975167,0.929768,0.958114,0.974029,0.975561,0.076645,0.904219,-0.155044,0.921675,0.681005,0.549442,0.056679,0.859245,0.752978,0.069373,0.624719,0.786817,0.619295,0.810363,0.588373,1,1,0.487852,0.746433,0.071418
36,-0.01536,-0.065609,-0.006744,-0.151705,-0.01355,-0.003424,0.009998,-0.015284,-0.012075,-0.015143,-0.211846,-0.222402,0.008821,0.507749,0.496464,0.472384,0.507257,0.505589,0.052695,0.460376,-0.196059,0.479462,0.379317,0.322967,-0.197464,0.484788,0.535293,0.066835,0.409872,0.461163,0.385183,0.448213,0.363722,0.487852,0.487852,1,0.500001,0.08076
37,-0.005263,-0.021884,-0.006549,-0.218591,-0.004029,0.000505,0.012559,-0.004946,-0.005468,-0.006145,-0.305124,-0.308339,0.00991,0.759049,0.726711,0.777732,0.753193,0.761349,-0.042211,0.664939,-0.059611,0.699368,0.511363,0.42937,0.08545,0.664343,0.560471,-0.152255,0.311463,0.565838,0.271888,0.554328,0.238706,0.746433,0.746433,0.500001,1,0.052993
38,-0.145099,-0.045935,-0.069497,-0.01768,-0.161048,-0.014082,0.003141,0.079689,-0.127691,-0.15853,-0.131942,-0.136032,-0.039037,0.082222,0.097781,0.06402,0.080757,0.074756,0.007703,0.304604,-0.009963,0.087176,0.060575,0.699171,-0.143793,0.338659,0.250467,0.299463,0.065803,0.072468,0.047944,0.064992,0.200552,0.071418,0.071418,0.08076,0.052993,1
}
\end{adjustbox}
\caption{Feature correlation heat map}\label{fig:heatmap}
\end{figure}